\documentclass[a4paper,twoside, twocolumn]{cpc-hepnp}
\usepackage[english]{babel}
\usepackage{lineno,hyperref, xspace,fancyhdr}
\modulolinenumbers[5]
\usepackage{makeidx}
\usepackage{color}
\usepackage{listings}
\usepackage{bm}
\usepackage{url}
\usepackage{booktabs}
\usepackage{amsmath}
\usepackage{paralist}
\usepackage{amssymb}
\usepackage{algorithm}
\usepackage{algorithmic}
\usepackage{upgreek}
\usepackage{textcomp}
\usepackage{multicol}
\usepackage{graphicx}
\usepackage{CJKutf8}

\newcommand{\kgyr}{kg$\cdot$yr\xspace}

\begin{document}
\begin{CJK*}{UTF8}{gbsn}



\title{Searching for Neutrino-less Double Beta Decay of $^{136}$Xe with PandaX-II Liquid Xenon Detector
\thanks{Supported by grants from the Ministry of Science and Technology of China
(No. 2016YFA0400301 and 2016YFA0400302), a Double Top-class grant from Shanghai
Jiao Tong University, grants from National Science Foundation of
China (Nos. 11435008, 11505112, 11525522, 11775142 and 11755001), grants from the Office of Science and Technology, Shanghai Municipal Government (Nos. 11DZ2260700, 16DZ2260200, and 18JC1410200), and the support from the Key Laboratory for Particle Physics, Astrophysics and Cosmology, Ministry of Education.
This work is supported in part by the Chinese Academy of Sciences Center for Excellence in Particle Physics (CCEPP) and Hongwen Foundation in Hong Kong.
Finally, we thank the following organizations for indispensable logistics and other supports: the CJPL administration and the Yalong River Hydropower Development Company Ltd.}
}%

\maketitle

\author{%
	Kaixiang Ni(倪恺翔)$^{1}$
	\quad Yihui Lai(赖奕辉)$^{1}$
  \quad Abdusalam Abdukerim(阿布都沙拉木·阿布都克力木)$^{1}$
  \quad Wei Chen(陈葳)$^{1}$
  \quad Xun Chen(谌勋)$^{1}$
	\quad Yunhua Chen(陈云华)$^{2}$
  \quad Xiangyi Cui(崔祥仪)$^{1}$
  \quad Yingjie Fan(樊英杰)$^{3}$
  \quad Deqing Fang(方德清)$^{4}$
  \quad Changbo Fu(符长波)$^{1}$
  \quad Lisheng Geng(耿立升)$^{5}$
  \quad Karl Giboni$^{1}$
  \quad Franco Giuliani$^{1}$
  \quad Linhui Gu(顾琳慧)$^{1}$
	\quad Xuyuan Guo(郭绪元)$^{2}$
  \quad Ke Han(韩柯)$^{1,\dagger}$\email{ke.han@sjtu.edu.cn}
  \quad Changda He(何昶达)$^{1}$
  \quad Di Huang(黄迪)$^{1}$
  \quad Yan Huang(黄焱)$^{2}$
  \quad Yanlin Huang(黄彦霖)$^{6}$
  \quad Zhou Huang(黄周)$^{1}$
  \quad Peng Ji(吉鹏)$^{3}$
  \quad Xiangdong Ji(季向东)$^{1,7}$
  \quad Yonglin Ju(巨永林)$^{8}$
  \quad Kun Liang(梁昆)$^{1}$
  \quad Huaxuan Liu(刘华萱)$^{8}$
  \quad Jianglai Liu(刘江来)$^{1,7,\ddagger}$
  \quad Wenbo Ma(马文博)$^{1}$
  \quad Yugang Ma(马余刚)$^{4}$
  \quad Yajun Mao(冒亚军)$^{9}$
  \quad Yue Meng(孟月)$^{1}$
  \quad Parinya Namwongsa$^{1}$
	\quad Jinhua Ning(宁金华)$^{2}$
  \quad Xuyang Ning(宁旭阳)$^{1}$
  \quad Xiangxiang Ren(任祥祥)$^{1,8}$
  \quad Changsong Shang(商长松)$^{2}$
  \quad Lin Si(司琳)$^{1}$
  \quad Andi Tan(谈安迪)$^{10}$
  \quad Anqing Wang(王安庆)$^{11}$
  \quad Hongwei Wang(王宏伟)$^{4}$
  \quad Meng Wang(王萌)$^{11}$
  \quad Qiuhong Wang(王秋红)$^{4}$
  \quad Siguang Wang(王思广)$^{9}$
  \quad Xiuli Wang(王秀丽)$^{8}$
  \quad Zhou Wang(王舟)$^{10,12,1}$
  \quad Mengmeng Wu(武蒙蒙)$^{3}$
  \quad Shiyong Wu(吴世勇)$^{2}$
  \quad Jingkai Xia(夏经铠)$^{1}$
  \quad Mengjiao Xiao(肖梦姣)$^{10,12}$
  \quad Pengwei Xie(谢鹏伟)$^{7}$
  \quad Binbin Yan(燕斌斌)$^{11}$
  \quad Jijun Yang(杨继军)$^{1}$
  \quad Yong Yang(杨勇)$^{1}$
  \quad Chunxu Yu(喻纯旭)$^{3}$
  \quad Jumin Yuan(袁鞠敏)$^{11}$
  \quad Dan Zhang(张丹)$^{10}$
  \quad Hongguang Zhang(张宏光)$^{1}$
  \quad Tao Zhang(张涛)$^{1}$
  \quad Li Zhao(赵力)$^{1}$
  \quad Qibin Zheng(郑其斌)$^{6}$
	\quad Jifang Zhou(周济芳)$^{2}$
  \quad Ning Zhou(周宁)$^{1}$
  \quad Xiaopeng Zhou(周小朋)$^{5}$
}

\centering {(PandaX-II Collaboration)}

\address{%
	~\\
	$^1$ INPAC and School of Physics and Astronomy, Shanghai Jiao Tong University, \\Shanghai Laboratory for Particle Physics and Cosmology, Shanghai 200240, China\\
	$^2$ Yalong River Hydropower Development Company, Ltd., 288 Shuanglin Road, Chengdu 610051, China\\
  $^3$ School of Physics, Nankai University, Tianjin 300071, China\\
  $^4$ Shanghai Institute of Applied Physics, Chinese Academy of Sciences, Shanghai 201800, China\\
  $^5$ School of Physics \& International Research Center for Nuclei and Particles in the Cosmos \&\\Beijing Key Laboratory of Advanced Nuclear Materials and Physics, Beihang University, Beijing 100191, China\\
  $^6$School of Medical Instrument and Food Engineering, University of Shanghai for Science and Technology, Shanghai 200093, China\\
  $^7$Tsung-Dao Lee Institute, Shanghai 200240, China\\
  $^8$School of Mechanical Engineering, Shanghai Jiao Tong University, Shanghai 200240, China \\
  $^9$School of Physics, Peking University, Beijing 100871,China \\
  $^{10}$Department of Physics, University of Maryland, College Park, Maryland 20742, USA \\
  $^{11}$School of Physics and Key Laboratory of Particle Physics and Particle Irradiation (MOE), Shandong University, Jinan 250100, China\\
  $^{12}$Center of High Energy Physics, Peking University, Beijing 100871, China\\
}

\begin{abstract}
We report the Neutrino-less Double Beta Decay (NLDBD) search results from PandaX-II dual-phase liquid xenon time projection chamber.
The total live time used in this analysis is 403.1 days from June 2016 to August 2018.
With NLDBD-optimized event selection criteria, we obtain a fiducial mass of 219 kg of natural xenon.
The accumulated xenon exposure is 242~\kgyr, or equivalently 22.2~\kgyr of $^{136}$Xe exposure.
At the region around $^{136}$Xe decay Q-value of 2458~keV, the energy resolution of PandaX-II is 4.2\%.
We find no evidence of NLDBD in PandaX-II and establish a lower limit for decay half-life of 2.4 $ \times 10^{23} $ yr at the 90\% confidence level, which corresponds to an effective Majorana neutrino mass $m_{\beta \beta} <  (1.3 - 3.5)$~eV.
This is the first NLDBD result reported from a dual-phase xenon experiment.

\end{abstract}



\footnotetext[0]{$\dagger$ Corresponding author: ke.han@sjtu.edu.cn}%
\footnotetext[0]{$\ddagger$  Spokesperson: jianglai.liu@sjtu.edu.cn}%

\end{CJK*}

\begin{multicols}{2}

\section{Introduction}
Neutrino-less double beta decay (NLDBD) is a hypothetical decay process, during which two neutrons in a nucleus decay to protons and emit two electrons but no neutrinos.
This process is a direct evidence that neutrinos are their own anti-particles, i.e., Majorana Fermions.
It also violates lepton number conservation since the emitted electrons are not accompanied by corresponding electron anti-neutrinos.
NLDBD is clearly a physical process beyond the Standard Model (SM) of particle physics~\cite{PDG}.
The SM-allowed double beta decay process with two electron anti-neutrinos emitted has been observed in a dozen or so isotopes, but NLDBD has not been convincingly observed yet.

There is a global effort to search for NLDBD with various candidate isotopes using different experimental approaches.
For example, $^{76}$Ge is widely used in high purity germanium semiconductor detectors~\cite{Agostini:2018tnm, Aalseth:2017btx,wang2017first}, while $^{136}$Xe is usually the target of monolithic liquid detectors~\cite{gando2016search,Albert:2017owj}. CANDLES-III~\cite{umehara2006candles}, GERDA~\cite{Agostini:2018tnm},  CUORE~\cite{alduino2018first}, and KamLAND-Zen~\cite{gando2016search} experiments have recently established the most stringent half-life lower limit respectively for $^{40}$Ca, $^{76}$Ge, $^{130}$Te, and $^{136}$Xe decaying into the ground state of their daughters.
Those limits are on the order of $10^{25}$ - $10^{26}$ years and show an extremely rare nature of NLDBD, if exists at all.

The PandaX-II detector~\cite{Tan:2016diz} is a large dual-phase liquid xenon time projection chamber (TPC).
The primary science goal is to directly detect possible collisions of dark matter candidate particle WIMP (Weakly Interacting Massive Particle) and xenon nuclei~\cite{Tan:2016diz, tan2016dark,cui2017dark}.
However, with a target mass of 580\,kg of xenon and a natural isotopic abundance of 8.9\%, it contains 51.6\,kg of $^{136}$Xe and gives us the possibility to search for NLDBD with a reasonably large amount of isotope.

In this paper, we present the NLDBD search results with 403.1 days of data from the PandaX-II experiment.
Section~\ref{sec:detector} will describe the PandaX-II detector and the data taking campaigns in year 2016-2018.
In Section~\ref{sec:datarecon}, we outline the data analysis procedure, with the energy reconstruction and event selections adjusted specifically for the high energy region of interest (ROI).
In Section~\ref{sec:prefit} and~\ref{sec:results}, we establish the background model from Monte-Carlo and fit the energy spectrum to identity possible NLDBD signal.
We also discuss in details different contributions to systematic uncertainties.
We summarize and give an outlook for NLDBD physics potentials with large liquid xenon detectors in the last section.

\section{PandaX-II detector and data taking campaigns}\label{sec:detector}

Located at China Jin-Ping Underground Laboratory (CJPL)~\cite{wu2013measurement}, the PandaX-II experiment has been taking physics data since 2016.
The detector is a dual-phase xenon TPC~\cite{DualPhaseTPC}, as shown in Fig.~\ref{TPC-View}.
Detailed description of the detector can be seen in~\cite{Tan:2016diz}.
The sensitive volume is defined by a cylinder of polytetrafluoroethylene (PTFE) reflectors of inner diameter 646 mm and two sets of photomultiplier tube (PMT) arrays at the top and bottom.
The drift/extraction/amplification field in the TPC is exerted by the cathode, gate, and anode meshes, and is constrained by a set of copper shaping rings outside the reflector barrel.
Particle going through the liquid generates scintillation photons and ionized electrons, the latter of which drift to the gas phase and generate electroluminescence photons.
Light signals are used to reconstruct energy and position information of a particular event.
Vetoing volume resides outside of the PTFE reflectors and is used to reject environment and detector background events originating from outside.

Multiple layers of passive shielding outside the TPC include  copper, inner polyethylene,  lead, and  outer polyethylene (see Fig.~\ref{shielding}).
Bricks or panels of those materials are concatenated into octagonal shape.
The gap between the innermost layer and the TPC vessel is flushed by nitrogen gas in order to suppress radon concentration inside.
Ambient radiation such as gamma and neutron background, is attenuated effectively with the help of passive shielding.

\begin{center}
   \includegraphics[width=6.5cm]  {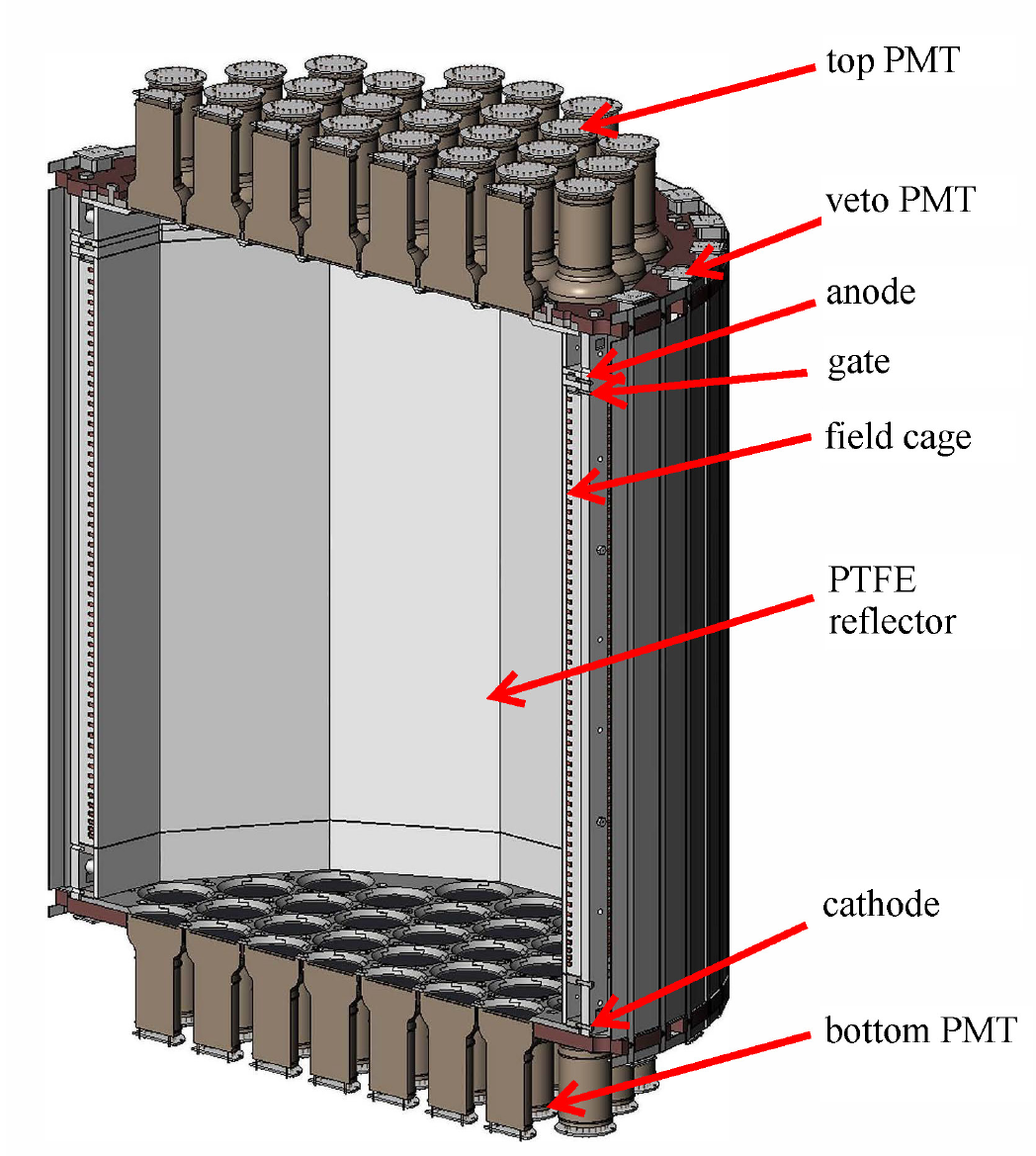}
	\figcaption{	\label{TPC-View}
Schematic drawing of the PandaX-II TPC.  Key components such as top and bottom PMT arrays, PTFE reflectors, and electrodes are highlighted. The figure is adapted from~\cite{Tan:2016diz}}.
\end{center}

\begin{center}
    \centering{\includegraphics[width=\columnwidth]  {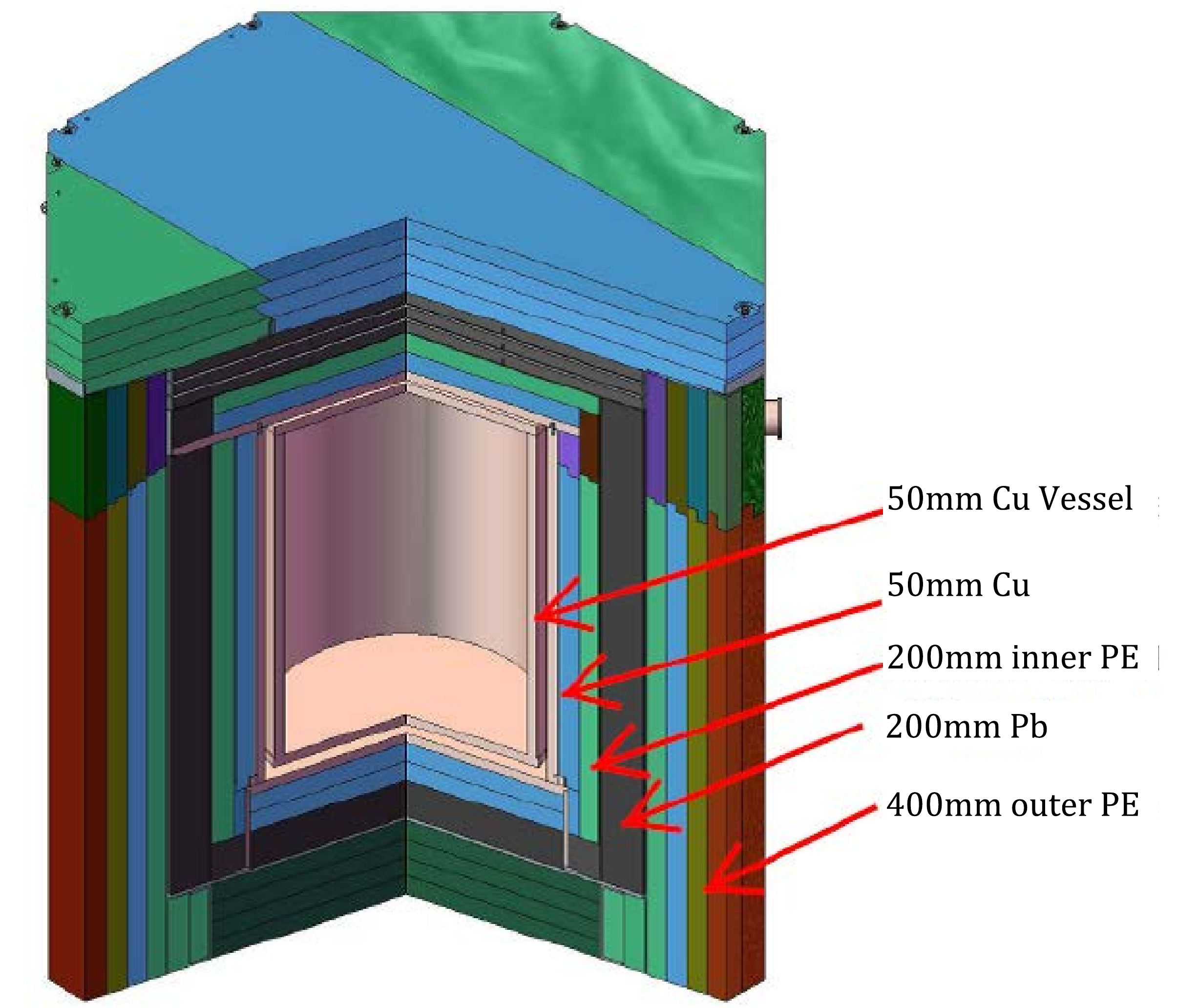}}
	\figcaption{\label{shielding}Schematic drawing  of the layered passive shielding system around the PandaX-II Detector.
Thickness of different layers are labelled.
The figure is adapted from~\cite{cao2014pandax}}.
	
\end{center}

In this paper, three sets of physics data from the past 3 years are used for NLDBD analysis, as shown in Table~\ref{tab:runs}.
The total accumulated exposure is 403.1 days.
Run numbers in the table follow the convention used in dark matter direct detection physics analysis.
Run 9 and 10 are the first two sets of physics runs for PandaX-II and have been previous published~\cite{tan2016dark,cui2017dark}.
Run 11 is a continuation of Run 10 with no change in run conditions.
Run 11 data taking was undertaken smoothly from July 2017 to August 2018, interrupted only by a series of calibration runs, as shown in Fig.~\ref{Exposure}.
The figure also shows that the drift electron lifetime, which is an indication for detector performance, was stable at above 400~\textmu s most of the time.
During this run, an additional 246.4 live-days of data was accumulated.
This is the first time that we present an analysis with this new data set.

\begin{center}
    \tabcaption{    \label{tab:runs}
PandaX-II data sets used for this NLDBD analysis.}
    \centering
    \begin{tabular*}{\columnwidth}{lccc}
    \toprule
    Data set & Begin & End & Live days \\
    \hline
    Run 9  & Mar. 9, 2016  & Jun. 30, 2016 & 79.6 d\\
    Run 10 & Apr. 22, 2017 & Jul. 16, 2017 & 77.1 d \\
    Run 11 & Jul. 17, 2017 & Aug. 16, 2018 & 246.4 d\\
    \bottomrule
    \end{tabular*}

    \end{center}
\begin{center}
      \centering{\includegraphics[width=\columnwidth]  {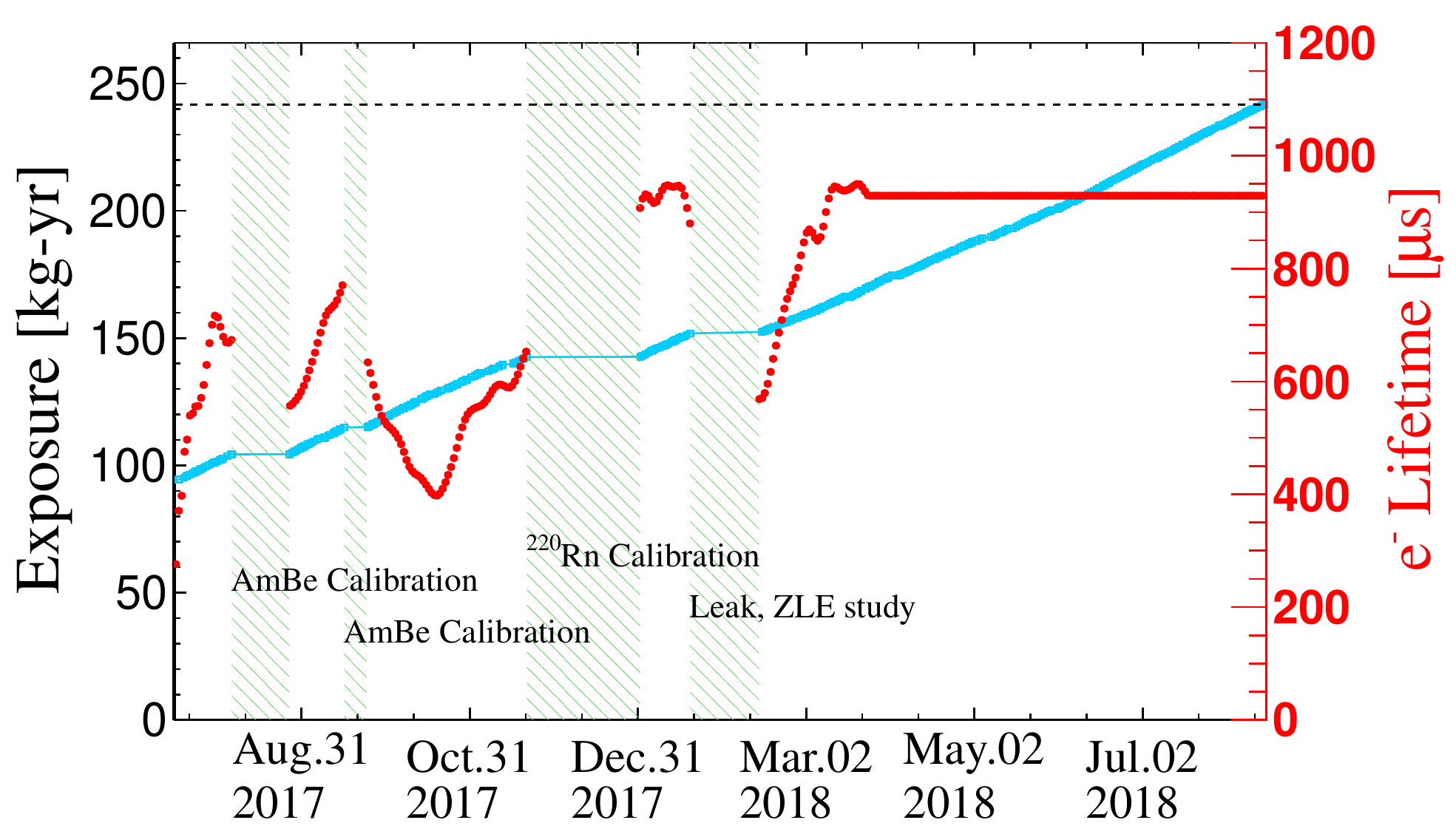}}
    \figcaption{\label{Exposure} Data taking history of Run 11.
    Other than 4 interruptions of calibration and detector performance study, we took data continuously from July 2017 to August 2018.
    The left Y axis is the exposure in the unit of \kgyr.
    The dash horizontal line indicates the total exposure at 242 \kgyr, with a corresponding fiducial mass of 219 kg.
    The right Y axis is the electron lifetime while drifting in the field cage.
}

\end{center}

\section{Energy reconstitution and event selection at MeV region}\label{sec:datarecon}

PandaX-II data analysis at MeV level follows the workflow of dark matter analysis closely~\cite{Xiao:2015psa}.
However, we developed new algorithms for energy reconstruction and new event selection criteria at MeV level specifically for NLDBD analysis.
In this section, we summarize the analysis workflow and highlight the key differences.

\subsection{PandaX-II data analysis workflow}
PandaX-II data acquisition system (DAQ) records triggered waveforms of prompt scintillation light signal (commonly called $S1$) and electroluminescence light signal ($S2$).
For each triggered signal, the readout window is 1~ms and the trigger point is placed in the middle of the signal window.
Data are recorded by underground DAQ computers and immediately copied to a local computing farm aboveground at CJPL, where first level data preprocessing is performed \emph{in situ} automatically.

The first step of data preprocessing chain process waveform of individual PMT.
After the Single Photoelectron (SPE) correction, which we will describe later, we group these waveforms into individual pulses, known as hits, using a hit finding algorithm.

The second step is to group coincident hits of PMTs to construct light signals.
Waveforms from individual pulses are summed up and the key parameters are calculated for the summed pulse.
We then classify these signals into $S1$, $S2$ or noise using a decision tree based on the characterization of summed waveforms.
A typical event may include one $S1$ signal and one or multiple $S2$ signals.
Each $S2$ signal paired with the preceding $S1$ signal corresponds to one interaction vertex (usually called \emph{site}) in the liquid xenon medium.

Finally we reconstruct the positions of sites in each event.
The $Z$ direction follows electron drift direction and points upward while $X$ and $Y$ directions define the horizontal plane.
The $X$ and $Y$ coordinates of each track are calculated from $S2$ signal.
Multiple methods are used in this step, including center of gravity (COG), template matching (TM), and maximum likelihood finding using photon acceptance function (PAF)~\cite{tan2016dark}, the last of which gives best uniformity of event distribution of calibration source, and is used as standard position reconstruction algorithm in the following analysis.
The $Z$ coordinate is calculated from electron drift velocity in the electric field and the drift time \emph{dt}, defined as time difference between $S1$ and $S2$ signals.
In Run~9 the drift velocity  is $1.71\pm0.01$ mm/\textmu s with an electric field of 393.5~V/cm.
In Run~10 and Run~11, the drift velocity dropped to $1.67\pm0.01$ mm/\textmu s due to the lowered drift field of 311.7~V/cm.
The derived drift velocities are consistent with Ref.~\cite{yoshino1976effect}.

\subsection{ Energy Calibration and Reconstruction}

The detector's response at MeV level is calibrated using external $^{232}$Th source capsules.
The source is deployed through the calibration loops inside the TPC outer vessel.
A full absorption gamma peak from $^{208}$Tl at 2615~keV can be seen in both $S1$ and $S2$ signals.
For MeV scale signals, a large $S2$ happens right below the top PMT array and the number of photons collected by a PMT may be beyond the linear response region.
In rare cases when signal size exceeds the dynamic range of digitizers, the waveform is also clipped.
Saturation of PMTs and digitizers deteriorates energy resolution.
A group of saturated pulses with both PMT and digitizer saturation is shown in Fig.~\ref{saturatedwf}.

\begin{center}
  \centering{\includegraphics[width=\columnwidth]  {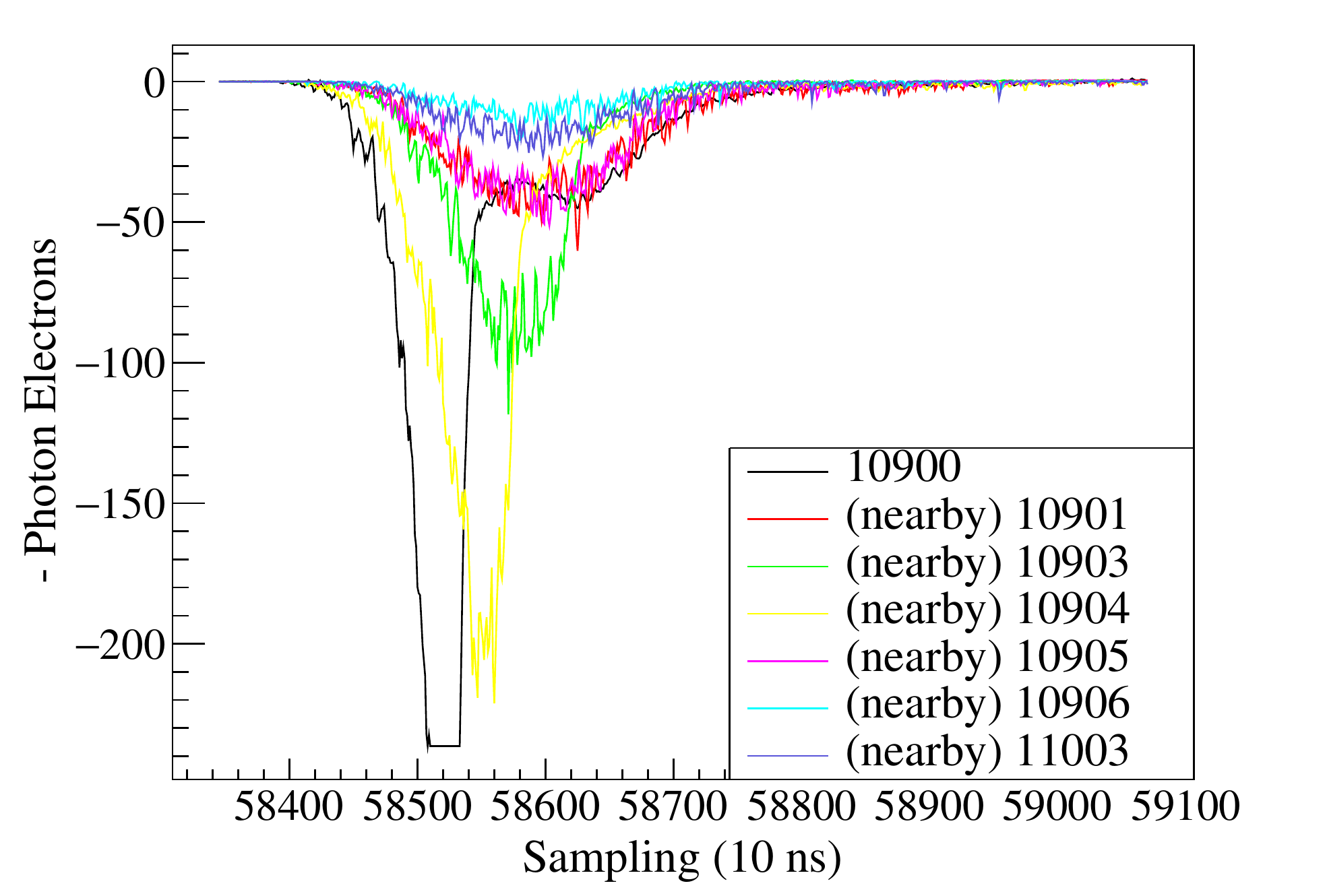}}
  \figcaption{\label{saturatedwf}
Examples of saturated waveform of PMTs.
  Different traces show waveforms from different top PMTs from a high energy event.
  The saturated PMT (black and yellow line) has asymmetric peak in its waveform, where the rising edge is steeper compared to the other ones.
  For PMT 10900, a typical PMT that experiences saturation problem, clipped peak is also observed as the digitizer saturates.
}
\end{center}

In our analysis, we used pulses from the bottom PMT array to reconstruct the energy of $S2$ signals to avoid the saturation problem.
Saturation is not observed in the bottom PMT array, since $S1$ signal is usually much smaller than $S2$ and $S2$ signal is far away from the bottom array.
We describe the whole process  below and emphasis is given on last two steps, where pulse saturation has key impact.

\begin{description}
  \item[SPE correction]
  Hit pulses are scaled  by the individual SPE values of PMTs so that the area of pulse represents the number of photoelectrons (PE).
  This correction compensates possible PMT gain drift during detector operation.
  \item[Position uniformity correction]
  The number of PE each PMT collects for any specific event depends on the $X$, $Y$, and $Z$ position of the interaction vertex.
  The position dependence is mapped out with internal radioactive sources such as  $^{83}$Kr or activated $^{127m}$Xe.
  $^{127m}$Xe emits 164~keV gamma rays and is better suited for position dependence correction at the high energy region.
  We divide the sensitive volume into small voxels to map out detector response at different position.
  In turn PMT response for any events is corrected with the mapping.

  Secondary electrons from xenon ionization are recombined with or attached to the xenon medium in the drifting process.
  The electron loss is usually characterized by electron life time in the unit of \textmu s.
  The measured $S2$ signal at liquid-gas interface is corrected with electron life time on a daily basis.

  \item[Energy reconstruction] To reconstruct the energy of an event, we follow the energy sharing formula used in dark matter analysis:
\[
E=W \times (\frac{S1}{PDE}+\frac{S2}{\textrm{EEE} \times \textrm{SEG}}).
\]
Here W equals to 13.7\,eV and  is the average work function for a particle to excite one electron-ion pair in xenon~\cite{Lenardo2015NEST}.
$S1$ is the total detected PE number from both top and bottom PMTs, after event position correction.
Since the saturated top PMT charge cannot be used, we replace $S2$ with bottom PMT PE number:
\[
S2=S2_{\textrm{Bottom}} \times (T/B+1),
\]
where $T/B$ stands for ratio between the numbers of PE from the top and bottom PMT array.
We expect that photon emission as well as collection is energy independent and set $T/B=2.3$, as derived in the dark matter analysis~\cite{tan2016dark}.
The other three parameters, photon detection efficiency (PDE), electron extraction efficiency (EEE), single electron gain (SEG), are derived by scanning the parameter space as follows.
For each combination of PDE and EEE $\times$ SEG, a spectrum of $^{232}$Th calibration source is obtained and fitted.
The combination of 7.4\% and 22.75 PE/e yields the best energy resolution and is selected.
The achieved  best energy resolution is 4.2\% at 2615~keV.
The PDE and EEE $\times$ SEG are different from those used in dark matter analysis, which implies changes of detector response in this different energy region.
The new set of values is verified with the anti-correlation of $S1$ and $S2$ bottom signals, as shown in Fig.~\ref{anti-correlation fit}.

\item[Residual non-linearity correction]
A spectrum with optimized energy resolution can be obtained with the above mentioned procedures.
However, peaks could be identified slightly off the true energy of the characteristic gamma lines.
Another correction is carried out to shift those peaks and correct the non-linearity of detector's energy response.
With a relatively large energy resolution, gamma lines on top of continuum background may not peak in the exact gamma energy.
Monte carlo simulation is used to study such shifts.
The detailed simulation process is discussed in Sec.~\ref{sec:prefit}.
Peaks from 915 to 2615~keV are used to correct the residual non-linearity, as shown in Fig.~\ref{energy_alignment}.
For each run, six peaks are used to fit a third-order polynomial function with zero intercept.
Then energy of each event is corrected with these fit functions.
\end{description}

\begin{center}
    \centering{\includegraphics[width=\columnwidth]  {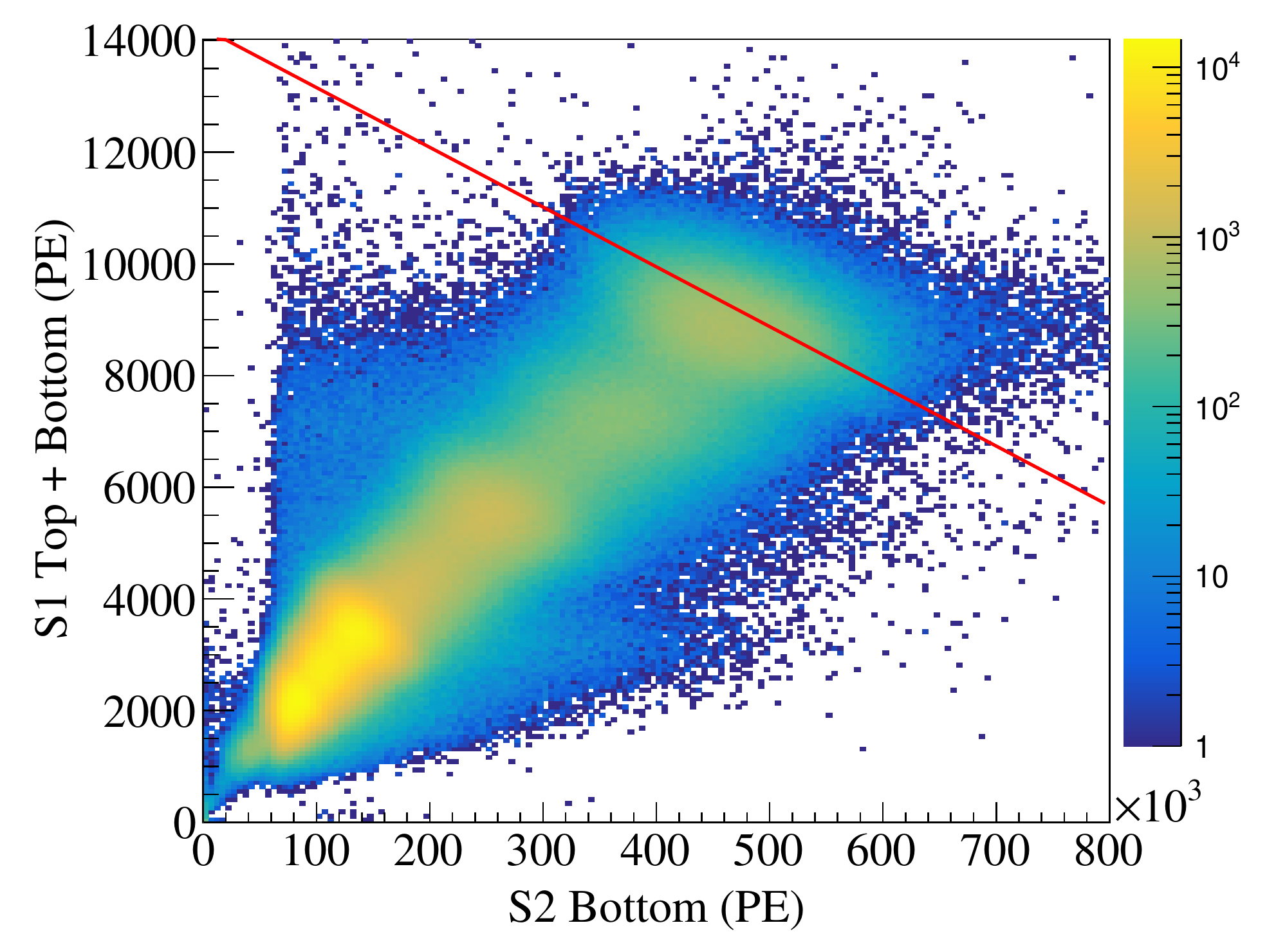}}
    \figcaption{	\label{anti-correlation fit}
Anti-correlation between $S1$ signals (Top + Bottom PMTs) and $S2$ Bottom PMT signals in $^{232}$Th calibration data.
The bright blob on the top-right is the 2615~keV gamma peak.
The red line indicates the best anti-correlation between $S1$ and $S2$ signals, whose slope is calculated from the ratio of PDE and EEE $\times$ SEG.
}
\end{center}

\begin{center}
    \centering{\includegraphics[width=\columnwidth]  {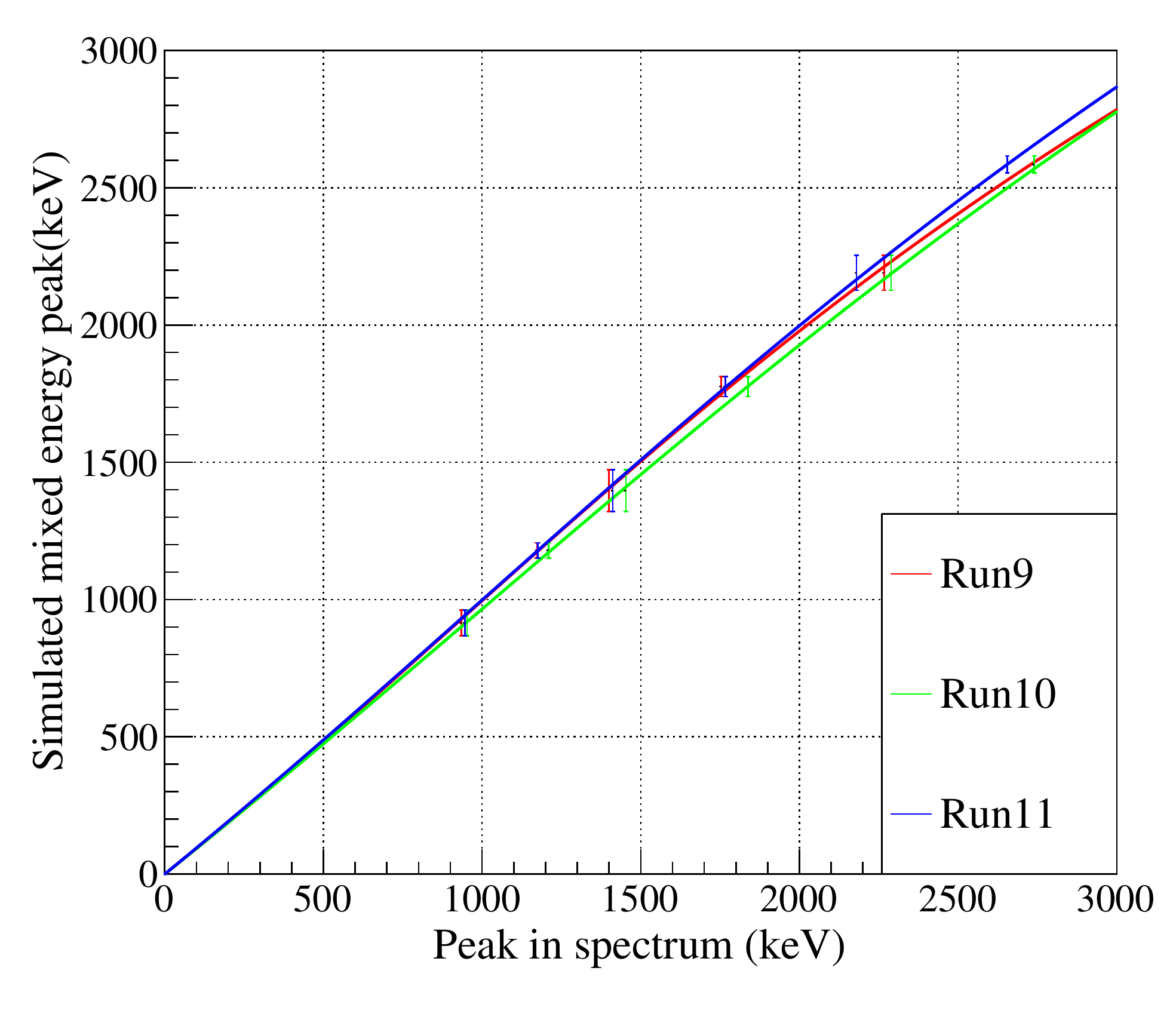}}
	\figcaption{	\label{energy_alignment}
Peak matching and residual nonlinearity of the three runs.
The X axis is the detected peak position from 915 to 2615~keV before corrections.
The Y axis is the expected peak position from Monte-Carlo simulation.
Three-order polynomial functions are used to fit the data to determine the correction functions for corrections.}

\end{center}

\subsection{Event Selection Cuts}

The original cuts from dark matter analysis are inherited but tuned to adapt to the high energy region in this study.
We first apply the single-site cut, ROI energy cut, veto cut, and fiducial volume cut to reject a vast majority of background events.
The signal quality cuts, including a few pulse shape cuts, $S1/S2$ band cut,  top/bottom asymmetry cut, and position quality cuts, are used to further reject possible low quality signals.
Here we briefly describe the cut criteria and more detailed definition can be found in~\cite{cao2014pandax,Xiao:2015psa}.

\begin{description}
\item[Single-site cut]
We focus on single-site events from NLDBD and reject all events with multiple $S2$ signals, because in most cases the two electrons emitted from an NLDBD event deposit energy in one point-like site in liquid xenon.
Monte Carlo shows that such selection accepts 93.4\% of all the NLDBD events, and the inefficiency is dominated by the multi-site deposition due to the Bremsstrahlung.
On the other hand, background events from gammas are predominately multi-site events, the cut reduces gamma background effectively.
For example, in Run 10, 6.9 million out of 18.4 million of triggered events are tagged as single-site events.

\item[ROI cut]
In the NLDBD final spectrum fit, we use events with energy in the range of 2058 to 2858 keV, i.e. $\pm400$~keV around the Q-value.
\item[Veto cuts]
The large majority of NLDBD events inside the field will not deposit any energy outside.
On the contrary, penetrating particles such as muon may deposit energy in the veto region and is rejected with an offline coincidence analysis between veto signal and $S1$ signal.
\item[Fiducial volume cut]
Fiducial volume for NLDBD analysis is optimized according to the distribution of high energy events.
Fig.~\ref{EventDist_High} shows the spatial distribution of high energy events in the ROI.
The X axis shows square of event radius.
The Y axis is the negative drift distance, which is equivalent to the relative vertical coordinate in space.
Number of events per bin remains relatively constant along the radius direction since xenon self-shielding is not as effective for high energy events.
At the edge of the sensitive volume, an abnormal decrease of event density is observed, which is due to malfunctioning of PAF position reconstruction algorithm with saturated PMT pulses at the detector's edge.
To be conservative, we force $r^2<80000\,\textrm{mm}^2$.
More background events concentrate at the top of the sensitive volume where the top PMT array, stainless steel flanges and pipes contribute significantly.
Therefore, the fiducial volume cut in vertical direction is not symmetric.
The vertical cut is within -233 and -533~mm,  optimized by comparing fiducial mass and the number of background events under different ranges.
The final fiducial region is shown in the red rectangle in Fig.~\ref{EventDist_High} and the fiducial mass within is 219~kg.
The fiducial volume cut is the strongest cut in the event selection, reducing large amount of background events by the self-shielding effect of liquid xenon.

We estimate the uncertainty of the fiducial mass with position reconstruction uncertainties.
The uncertainty of position reconstruction in the radial direction is 21.5~mm and given by the mean difference between of PAF and TM method for events in the ROI.
The uncertainty in the vertical direction is 3~mm, mainly from the uncertainty of drift velocity measurement.
Overall, we obtained an uncertainty of 32~kg for our fiducial mass.
\begin{center}
    \includegraphics[width=0.9\columnwidth]  {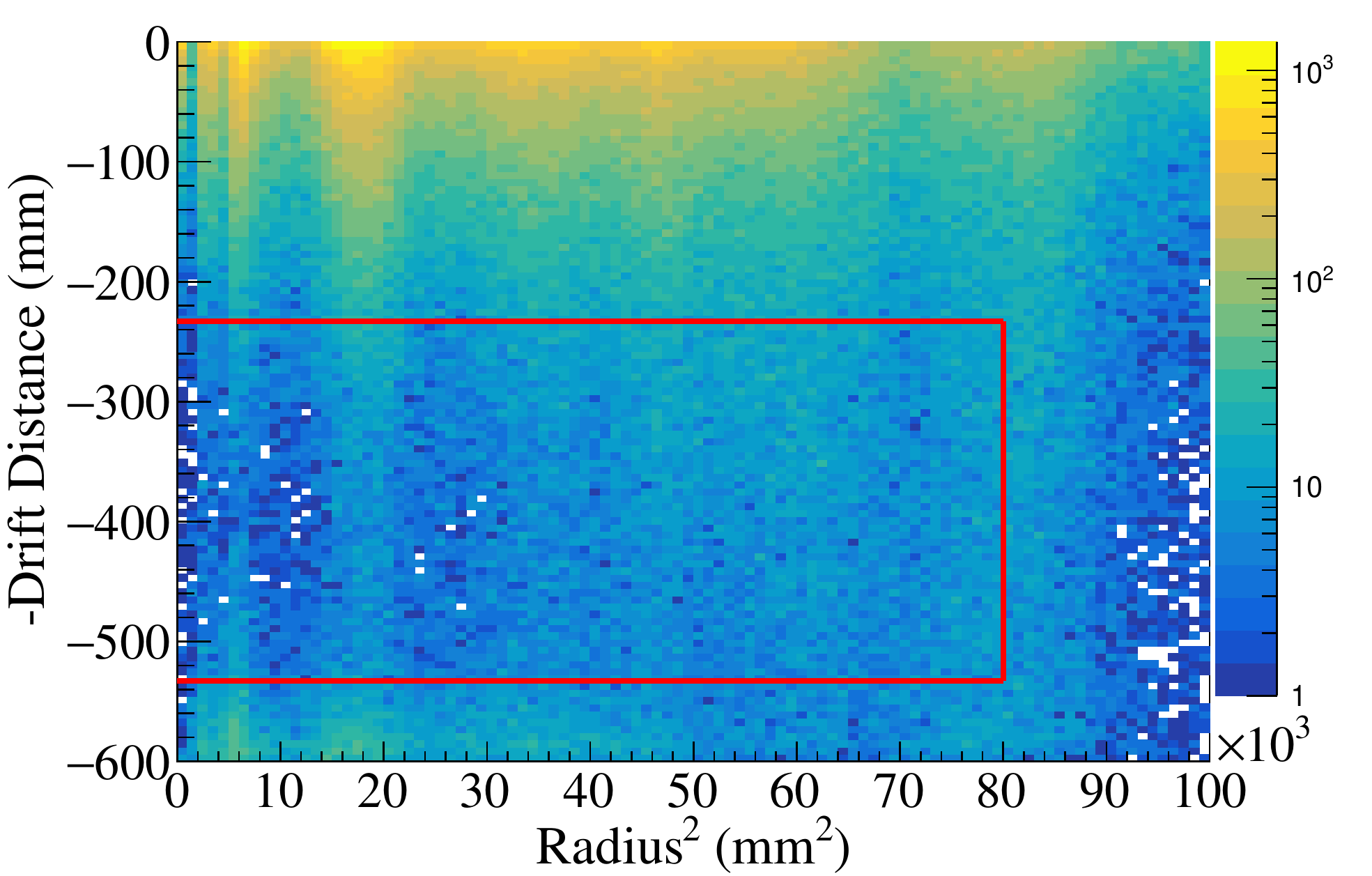}
	\figcaption{	\label{EventDist_High}
Spatial distribution of ROI event in the PandaX-II TPC.
The Y axis is in the electron drift direction with the liquid and gas interface at 0~mm.
Single scatter cut, quality cut, and veto cut are added.
Fiducial volume used in the NLDBD analysis is marked as red rectangle.
}
\end{center}
  \item[Pulse shape cuts]
 Two pulse shape cuts are used to reject pulses with positive overshoot.
\emph{Ripple ratio cut} calculates the overshoot of $S1$ waveforms and rejects pulses with a ratio larger than 0.5\%.
\emph{Negative charge cut} compares two methods of total PMT charge calculation.
A discrepancy between them also indicates unwanted overshoot in corresponding waveforms.
We reject events with such discrepancy larger than 1.5\%.
\item[$S1/S2$ band cut]
Noise signals of electrode discharge have pulses similar to $S2$ but with narrower width.
Consequently, events with \emph{small ``S3''} are rejected, whose log$_{10}(S2/S1)$ are smaller than 1.5.
\item[Top/bottom  asymmetry cut] The cut is defined as the difference between numbers of PE collected by top and bottom PMT arrays for $S1$ signals.
Random coincidence events, usually with large asymmetry, are rejected.
\item[Position quality cuts]
This group of cuts reject events with poor position reconstruction quality.
Events with unsatisafactory PAF fitting quality are rejected.
We also require the difference between positions reconstructed by PAF and TM methods to be less than 40~mm.

\end{description}

Impacts of these successive cuts are shown in Table~\ref{survival_events} and Fig.~\ref{full_spectrum_scaled}.
In Table \ref{survival_events} we show the number of survived events and estimated signal efficiencies of the relevant cuts.
Detector trigger efficiency at high energy region is estimated to be about 100\% and neglected here.
With a wide ROI, almost all the NLDBD events would fall in the region and the ROI cut efficiency is neglected.
Veto cut efficiency is not needed since we only study single-site NLDBD events with full energy deposition.
Fiducial volume cut is reflected in the final $^{136}$Xe mass calculation.
Efficiencies of signal quality cuts are evaluated conservatively by comparing the number of events before and after applying corresponding cut.
For example, the pulse shape cut efficiency is $99.3\pm0.8$\%, averaged over the values of three runs.
By combining all the cuts, we calculate NLDBD detection efficiency to be $91.6 \pm 0.8$\%.
Meanwhile, the number of background events has been suppressed by three orders of magnitude.

\begin{center}

    \tabcaption{    \label{survival_events}
Event selection tree and the efficiency for NLDBD signal.
    Signal efficiency stands for the ratio of NLDBD events that can pass through the step.
        }
    \begin{tabular*}{\columnwidth}{lccc}
    \toprule
                                & Events        & Signal efficiency (\%) \\
    \hline
    Single scattering cut       & 38070871       & 93.4            \\
    ROI cut                     & 771356       & $\approx$100             \\
    Veto cut                    & 433482       & -             \\
    Fiducial volume cut         & 32132        & -             \\
    Pulse shape cut             & 31900        & 99.3 $\pm$ 0.8 \\
    $S1/S2$ pattern cut         & 31837        & 99.8 $\pm$ 0.1 \\
    Position quality cut        & 31511        & 99.0 $\pm$ 0.1 \\
    \hline
    Total                       & -            & 91.6 $\pm$ 0.8 \\
    \bottomrule
    \end{tabular*}

\end{center}

\begin{center}
\centering{\includegraphics[width=\columnwidth]  {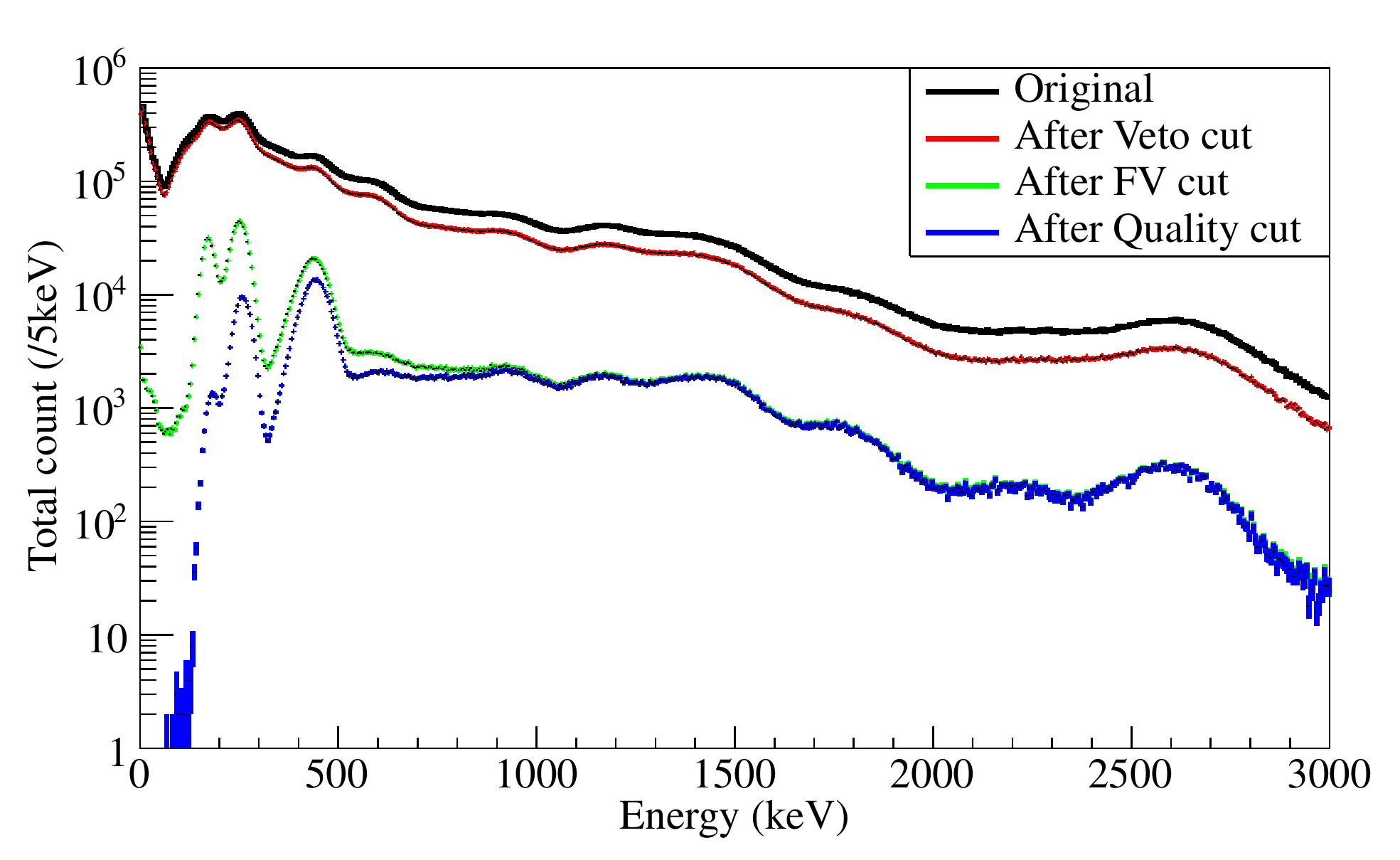}}
    \figcaption{\label{full_spectrum_scaled}Energy spectrum from 0 to 3~MeV after various cuts.
Impacts of different cut can be seen in the plot.
Backgrounds are rejected effectively by veto and fiducial volume(FV) cuts.
The signal quality cut is tuned specifically for high energy events, and yields a lower efficiency for low energy events far away from the ROI.}	
\end{center}

\section{Monte-Carlo simulation and background estimation}\label{sec:prefit}

Gamma events from $^{238}$U, $^{232}$Th, $^{222}$Rn, and $^{60}$Co originating from detector and xenon medium contribute most to the background in the NLDBD ROI.
Spectra from $^{222}$Rn and $^{60}$Co do not have any obvious peak in the ROI and can not be constrained well with fitting.
The background rates are determined with a Q-value-blinded pre-fit with a range between 1 and 3~MeV, excluding data within the $2458\pm100$~keV signal region.
In the extended range, we include the 1173 and 1332~keV peaks from $^{60}$Co, as well as additional peak from $^{40}$K, $^{238}$U, and $^{232}$Th.

Background spectral shape in the ROI from various components is estimated with Monte-Carlo simulation based on the Geant4 toolkit~\cite{agostinelli2003geant4}.
For a given radioactive source originating from a specific detector component, we simulate a large number of events and obtain the energy deposition information in xenon.
Detector response is also considered in the simulation.
Hits are merged together when their distance in Z direction is smaller than 8.5~mm, which corresponds roughly to the width of a $S2$ defined by the gas gap, below which we could not identify a double scatter event.
We also remove the events with multiple unmerged hits or with hits inside veto volume or outside fiducial volume.
Background contribution from $^{222}$Rn dissolved in the xenon liquid is out of secular equilibrium and considered separately.
The $^{214}$Bi in the $^{222}$Rn decay chain, a beta emitter with the associated 2447 keV gamma ray, is largely removed due to the delayed coincidence of $^{214}$Po (see later).
Its contribution from the veto region (about 100 kg of liquid xenon) can also be neglected by the active veto of beta energy and the self-shielding effects of liquid xenon.
For other isotopes, such as $^{238}$U, spectra from all the components are summed up with a weighting factor, which is derived from the relative radioactivity of each component, as reported in Ref.~\cite{Wang2016counting}.

In addition, we take the two neutrino double beta decay (2NDBD) of $^{136}$Xe into consideration.
Energy distribution of two emitted electrons from 2NDBD is simulated with the decay0 generator~\cite{ponkratenko2000event}, and then the summed electron energy spectrum is added to the fit.
The 2NDBD event rate is fixed to 48 per \kgyr in the fit region, which is calculated according to $^{136}$Xe's 2NDBD half-life~\cite{Albert:2013gpz} and its abundance in natural xenon.

The fit is done with a software package called Bayesian Analysis Toolkit (BAT)~\cite{caldwell2009bat}.
This software applies Bayes' Theorem to optimize multiple variables when given a likelihood function.
The fit is done for each run independently, and the summed fit result is shown in Fig.~\ref{spectrum_fit}.
The fitted isotope rates are propagated to the final NLDBD fit, used as inputs for Gaussian priors.
Four parameters are added in the fit model to describe energy resolution at 1173 ($^{60}$Co), 1447 ($^{40}$K), 1764 ($^{214}$Bi), and 2615 ($^{208}$Tl)~keV position in the spectrum.
The energy resolution at any other energy is taken as the interpolation of resolutions of two nearby peaks.
Fitted energy resolutions at 2615~keV peak are $4.8\pm0.1\%$, $4.2\pm0.1\%$, and $4.3\pm0.1\%$ respectively.
Energy resolution in Run 9 is slightly worse, since the uniformity correction map with the 164~keV gamma peak was obtained with a calibration period only towards the end of that run.

\begin{center}
\includegraphics[width=\columnwidth]  {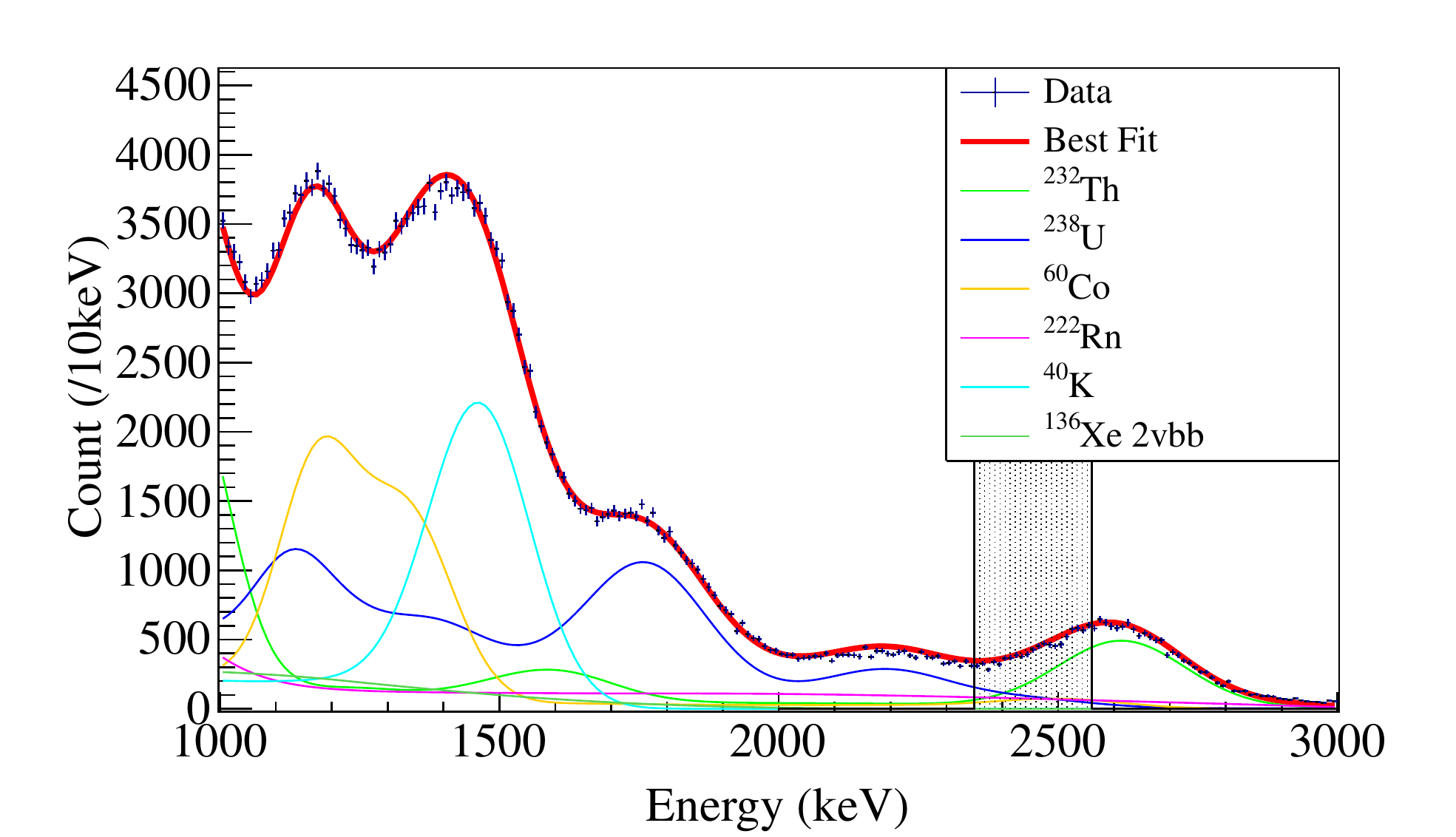}
    \figcaption{	\label{spectrum_fit}
Data and best-fit curves between 1 to 3~MeV to evaluate the background contributions from radioactive sources.
The spectrum is a summation of three runs.
The shaded region around NLDBD Q-value is not used in the fitting process.}
\end{center}

Background rate from radon in xenon medium is cross-checked with a so-called Bi-Po coincidence study.
For example, $^{214}$Bi, which is after $^{222}$Rn in the $^{238}$U decay chain, beta decays to polonium nucleus, with a series of signature gamma rays emitted.
$^{214}$Po quickly decays to $^{210}$Pb and a 7.7~MeV alpha particle with a half-life of 164.3~\textmu s.
Consequently, decay of $^{214}$Bi and $^{214}$Po is very likely to happen within a readout window width of 1~ms , which is easy to identify.
As discussed in Ref.~\cite{li2017krypton}, we select the sequential Bi-Po decay events by finding two adjacent $S1$ signals with time difference consistent with the polonium decay.
The concentration of $^{222}$Rn is measured to be (6.6 - 8.6) \textmu Bq/kg during the three runs, while $^{220}$Rn is an order of magnitude lower and ignored in the following analysis.
The statistical uncertainty of radon concentration is less than 1\% and neglected.
The relative systematic uncertainty of $^{222}$Rn is about 50\%, estimated from the discrepancy of $^{214}$Po single event rate and $^{214}$Bi-$^{214}$Po coincidence event rate~\cite{cui2017dark}.
We see a reduction of $^{222}$Rn concentration from Run 9 to 11, possibly due to changes in the circulation and purification conditions.
The results agree with our pre-fit results within $1\sigma$ uncertainties.

\section{Double beta decay fit results}\label{sec:results}

The background model of NLDBD ROI differs from the background pre-fit as described in the previous section in a few key aspects.
In the ROI fit  between 2058 to 2858 keV, energy resolutions are fixed to their values obtained in the pre-fit.
We ignore the negligible $^{40}$K and 2NDBD contributions in this fit.
Event rates from other isotopes are still to be fitted, but assigned with Gaussian priors with their sigma from the pre-fit results.

The fitted NLDBD rate $\Gamma_{0 \nu}$ is in unit count per year per nucleus.
To add it into model, we convert the decay rate to total count $C$:
\[
C = \frac{\Gamma_{0 \nu} \times T \times N_A \times R \times M \times \epsilon }{m} ,
\]
where $T$ is the detector live time, $N_A$ is the Avogadro constant, $R=8.9\%$ is the natural isotopic abundance of $^{136}$Xe, $M$ is the xenon mass in our fiducial volume, $\epsilon$ is the detection efficiency, and $m$ is the molar mass of xenon.

The best fit model combines data of three runs in one BAT fit.
Background rate is independent in each of the runs, while the NLDBD decay rate is forced to be one parameter.
In total we have $3\times4+1=13$ (3 runs, 4 background components) fit parameters for the model.
The fitted ROI spectrum is shown in Fig.~\ref{best_fit}.
Fit results of all free parameters are listed in Table~\ref{table_bestfit}.
The fitted NLDBD rate is $(-0.25\pm0.21) \times 10 ^{-23}\,\mathrm{yr}^{-1}$, with an equivalent event rate $-0.93\pm0.79$ per \kgyr

\begin{center}
\includegraphics[width=\columnwidth]  {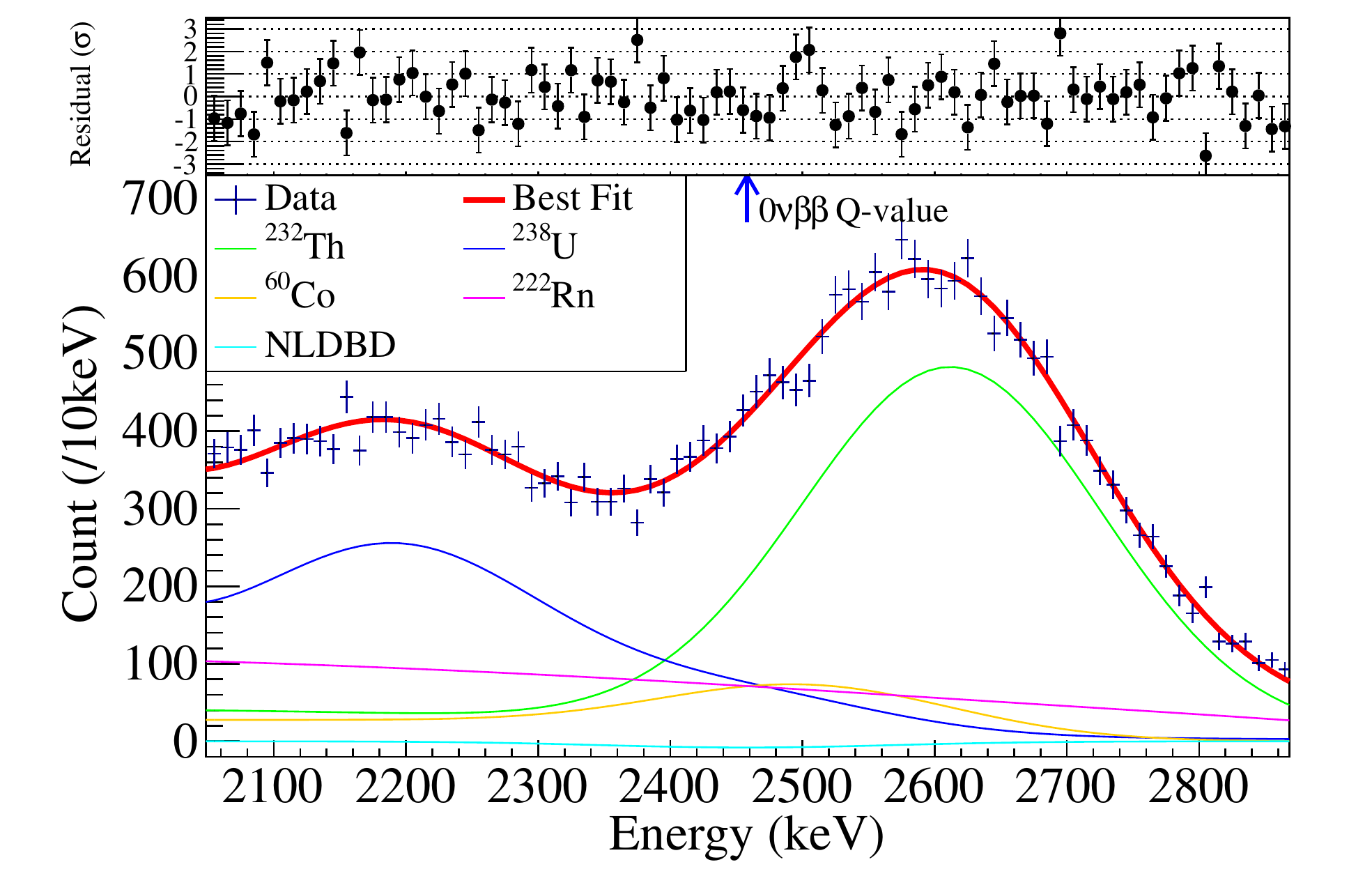}
	\figcaption{	\label{best_fit}
(Bottom) Energy spectrum and best-fit model for the NLDBD ROI.
Three runs are fitted simultaneously with shared NLDBD rates and the spectrum is the summation of all the data.
The fits are dominated by 2204 and 2615~keV gamma peaks from $^{238}$U and $^{232}$Th decay chain respectively.
An NLDBD rate of $-0.25\pm0.21 \times 10 ^{-23}\,\mathrm{yr}^{-1}$ is obtained from our best fit model.
(Top) The normalized residuals of the fits.}
\end{center}

\begin{center}
    \tabcaption{    \label{table_bestfit}
Summary of best-fit parameters.
    Event rates are in the unit of counts per \kgyr.}
    \centering
    \begin{tabular*}{\columnwidth}{lcccc}
    \toprule
    Item                 & Run 9           & Run 10           & Run 11           \\
    \hline
    $^{60}$Co            & 14.5 $\pm$ 1.0  & 13.3 $\pm$ 1.0   & 11.6 $\pm$ 0.6   \\
    $^{222}$Rn chain     & 33.4 $\pm$ 6.3  & 31.9 $\pm$ 5.2   & 22.9 $\pm$ 3.3   \\
    $^{232}$Th chain     & 63.6 $\pm$ 2.4  & 55.8 $\pm$ 2.0   & 67.6 $\pm$ 1.3   \\
    $^{238}$U chain      & 36.6 $\pm$ 4.2  & 34.8 $\pm$ 3.4   & 41.0 $\pm$ 2.1   \\
    \hline
    NLDBD                & \multicolumn{3}{c}{$-0.93 \pm 0.79$}                  \\
    \bottomrule
    \end{tabular*}

\end{center}

We also estimate systematic uncertainties induced by different parameters, as listed in Table~\ref{table_sys}.
As in Ref.~\cite{alduino2016analysis}, we split the uncertainty contribution into additive part ($\sigma_{\mathrm{add}}$, which is independent of the decay rate) and scaling part ($\sigma_{\mathrm{scaling}}$, which contributes to an uncertainty proportional to the true decay rate).
Uncertainties from the signal detection efficiency and fiducial mass only have the scaling part, and are first added to the table.

We carry out a set of pseudo-experiments with toy Monte-Carlo spectrum.
Toy spectra are generated according to the fitted background model and mixed with varied NLDBD events.
For example, for energy resolution $\Sigma$, we modify its value by its uncertainty in the background model to be $\tilde{\Sigma}$.
Hypothetical NLDBD decay rate $\tilde{\Gamma}$ is changed from 0 to $3\times10^{-23} \,\mathrm{yr}^{-1}$.
For each fake NLDBD rate $\tilde{\Gamma}$ we generate 1000 toy Monte-Carlo spectra according to the fitted background model but with $\tilde{\Sigma}$.
Then the spectra are re-fitted by imposing the best fit background model with all systematic terms fixed at their best fit values (e.g. $\Sigma$), yielding an NLDBD event rate which is biased from the input.
The fitted NLDBD rates are plotted against the input ones and the distribution is fitted with a linear function.
Interception of the fit on the Y axis represents the $\sigma_{\mathrm{add}}$ and its slope's deviation from one is $\sigma_{\mathrm{scaling}}$.

The energy scale at 2458 keV, estimated to be 96+-1\% (see Fig. 6), also contributes to the uncertainty of the NLDBD rate.
We modify its value by one sigma when producing the pseudo-experiment data, and repeat the constrained fits mentioned above.
The uncertainty of the fitting process itself, known as fit bias, is derived similarly but with no change of the systematic sources.

Results of systematic uncertainty evaluation are listed in Table~\ref{table_sys}.
Under the assumption that all the uncertainties are un-correlated, the total systematic uncertainty can be calculated as:
\[
\sigma_{\text { syst }}\left(\Gamma_{0 v}\right)=\sqrt{\sum_{i} \left(\sigma_{\mathrm{add}, i}^{2}+\sigma_{\mathrm{scaling}, i}^{2} \Gamma^{2}_{0 v}\right)}.
\]

\begin{center}
    \tabcaption{    \label{table_sys}
Summary of the additive and scaling systematic uncertainties of different contributions.
    The unit of additive systematic contribution is ($10^{-23}yr^{-1}$).
    The unit of scaling systematic contribution is (\%).
    Energy resolution and energy scale vary in the three runs, therefore only approximate values are shown.
    }
    \centering
    \begin{tabular*}{\columnwidth}{lcccc|}
    \toprule
                      & Fixed value                  & Additive         & Scaling  \\
    \hline
    Signal detection  & 91.6 $\pm$ 0.8 (\%)          &                  & 0.83   \\
    Fiducial mass     & 219 $\pm$ 32 (kg)            &                  & 14.7   \\
    Energy scale      & 96 $\pm$ 1 (\%)              &     0.045        & -   \\
    Energy reso.      & 4.2 $\pm$ 0.1 (\%)           &     0.089        & 2.47   \\
    Fit bias          & -                            &     0.041        & 0.04  \\
    \bottomrule
    \end{tabular*}

\end{center}

Our final NLDBD rate is $ (-0.25 \pm 0.21\,(\mathrm{stat.}) \pm 0.11\, (\mathrm{sys.})) \times 10^{-23}\,\mathrm{yr}^{-1}$.
The result is consistent with zero and we find no evidence of NLDBD in the PandaX-II data.
We can calculate a physical Bayesian limit from this best-fit NLDBD rate.
The posterior probability distribution and the systematic uncertainties are combined, as shown in Fig.~\ref{NLDBD dist}.
By integrating the positive part of the total likelihood curve to an area of 90\%~\cite{casella2002statistical}, we obtain a limit on NLDBD rate $\Gamma_{0 \nu}< 0.29 \times 10^{-23}$ yr$^{-1}$ and NLDBD half-life $T_{1/2}^{0\nu}> 2.4 \times 10^{23}$ yr at the 90\% confidence level (C. L.).
The half-life limit would be $2.9 \times 10^{23}$ yr if we only consider the statistical uncertainty.
Effective Majorana mass can be evaluated under the assumption of light Majorana exchange in the NLDBD process.
We adopt the phase space factors from~\cite{kotila2012phase} and the nuclear matrix elements calculated from models in~\cite{barea20150, rodriguez2010energy, engel2014chiral, menendez2009disassembling, mustonen2013large}.
The upper limit of effective Majorana neutrino mass $m_{\beta \beta}$ is found to be in the range of 1.3 and 3.5~eV at the 90\% C.~L..

\begin{center}
\includegraphics[width=\columnwidth]  {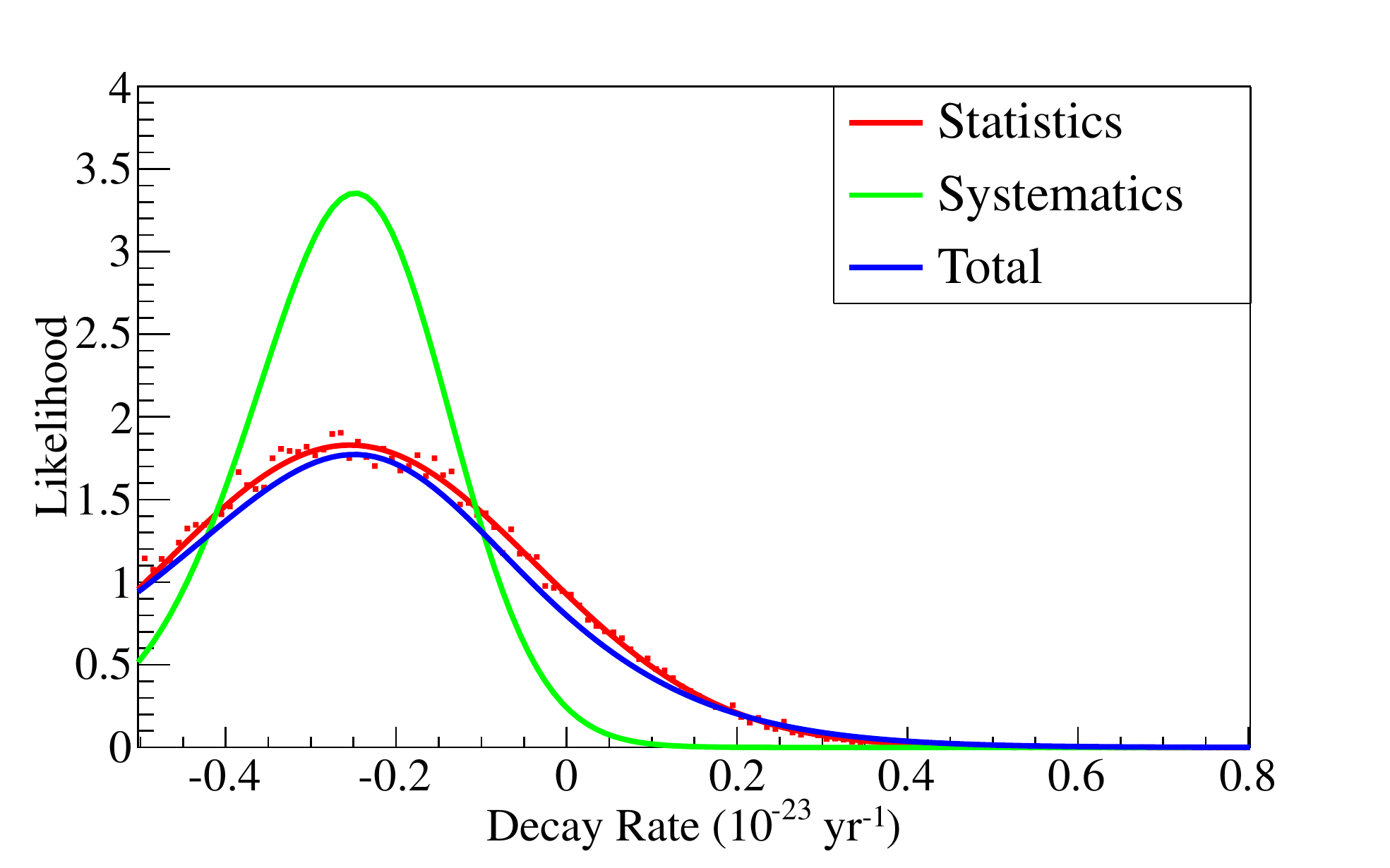}
	\figcaption{\label{NLDBD dist}NLDBD rate probability distribution.
The red scattered dots are from BAT fit result, while the red line is their gaussian fit.}
\end{center}

The sensitivity of $^{136}$Xe NLDBD search in PandaX-II experiment can also be calculated.
We carry out another set of pseudo-experiments, whose spectra are analyzed in the same way as our experimental data.
The fitted probability distribution curves are associated with the systematic table and NLDBD half-life limits at the 90\% C. L. are calculated.
A total of 12000 MC spectra are analyzed and the median value of derived half-life limits is chosen as the sensitivity.
The resulting NLDBD half-life sensitivity is $T_{1/2}^{0\nu}> 1.9 \times 10^{23} $~yr at the 90\% C. L..

\section{Summary and outlook}\label{sec:discussion}

We search for possible neutrino-less double beta decay signals from $^{136}$Xe with 244~\kgyr's data of PandaX-II detector from March 2016 to August 2018.
No evidence of NLDBD event is observed, and the half-life limit for $^{136}$Xe is given to be $T_{1/2}^{0\nu}>2.4 \times 10^{23} $ yr at the 90\% C. L. and the corresponding upper limit for $m_{\beta\beta}$ is the in range of 1.3 to 3.5~eV.
This work is the first NLDBD result from dual-phase liquid xenon detector, as far as we are aware.
We have developed new techniques of energy reconstruction and event selection, and studied systematically the detector performance in response of MeV-scale events at the PandaX-II detector.
The background budget of PandaX-II at high energy region is also presented for the first time in this analysis, and will be used to guide the design and construction of future detectors.

Due to limited performance in energy resolution and background rate, our result is not as competitive as the dedicated NLDBD experiments, such as KamLAND-Zen~\cite{gando2016search}.
However, it shows the multi-physics reach of dual-phase liquid xenon detector besides dark matter WIMP search.
PandaX-II has been focusing on signals and background in the keV range while NLDBD region of interest is mostly above 2~MeV, where PMT saturation issue deteriorates resolution significantly.
On the other hand, we recognize that the resolution is not intrinsically limited.
The XENON1T experiment has shown excellent energy resolution~\cite{XENON:2019dti}, comparable with other liquid xenon TPCs.
The PandaX collaboration is building the next generation PandaX-4T~\cite{zhang2019dark} detector and will use dual-dynode readout for PMTs, to avoid the PMT saturation issues altogether.
PandaX-4T is expected to have a much improved energy resolution for double beta decay physics.
PandaX-II also suffers from a high background rate in the NLDBD ROI since we can not take advantage of background suppression techniques developed for WIMP signals in the keV region.
Techniques such as nuclear recoil and electron recoil discrimination are specific to WIMP searches.
Effective self-shielding with detector medium at low energy become much less so at higher energy.
In the proposed 30-ton detector such as DARWIN~\cite{Baudis:2013qla} and PandaX future generations,  the fiducial volume in the core of the detector will be much more effectively shielded.
At this future stage, dual-phase liquid xenon detector with natural xenon will be able to explore the inverted neutrino mass hierarchy region in the $m_{\beta \beta}$ phase space.

\end{multicols}

\vspace{-1mm}
\centerline{\rule{80mm}{0.1pt}}
\vspace{2mm}

\begin{multicols}{2}

\end{multicols}


\begin{thebibliography}{34}%
    \makeatletter
    \providecommand \@ifxundefined [1]{%
     \@ifx{#1\undefined}
    }%
    \providecommand \@ifnum [1]{%
     \ifnum #1\expandafter \@firstoftwo
     \else \expandafter \@secondoftwo
     \fi
    }%
    \providecommand \@ifx [1]{%
     \ifx #1\expandafter \@firstoftwo
     \else \expandafter \@secondoftwo
     \fi
    }%
    \providecommand \natexlab [1]{#1}%
    \providecommand \enquote  [1]{``#1''}%
    \providecommand \bibnamefont  [1]{#1}%
    \providecommand \bibfnamefont [1]{#1}%
    \providecommand \citenamefont [1]{#1}%
    \providecommand \href@noop [0]{\@secondoftwo}%
    \providecommand \href [0]{\begingroup \@sanitize@url \@href}%
    \providecommand \@href[1]{\@@startlink{#1}\@@href}%
    \providecommand \@@href[1]{\endgroup#1\@@endlink}%
    \providecommand \@sanitize@url [0]{\catcode `\\12\catcode `\$12\catcode
      `\&12\catcode `\#12\catcode `\^12\catcode `\_12\catcode `\%12\relax}%
    \providecommand \@@startlink[1]{}%
    \providecommand \@@endlink[0]{}%
    \providecommand \url  [0]{\begingroup\@sanitize@url \@url }%
    \providecommand \@url [1]{\endgroup\@href {#1}{\urlprefix }}%
    \providecommand \urlprefix  [0]{URL }%
    \providecommand \Eprint [0]{\href }%
    \providecommand \doibase [0]{http://dx.doi.org/}%
    \providecommand \selectlanguage [0]{\@gobble}%
    \providecommand \bibinfo  [0]{\@secondoftwo}%
    \providecommand \bibfield  [0]{\@secondoftwo}%
    \providecommand \translation [1]{[#1]}%
    \providecommand \BibitemOpen [0]{}%
    \providecommand \bibitemStop [0]{}%
    \providecommand \bibitemNoStop [0]{.\EOS\space}%
    \providecommand \EOS [0]{\spacefactor3000\relax}%
    \providecommand \BibitemShut  [1]{\csname bibitem#1\endcsname}%
    \let\auto@bib@innerbib\@empty
    \bibitem [{\citenamefont {Olive}\ \emph {et~al.}(2014)\citenamefont {Olive}
      \emph {et~al.}}]{PDG}%
      \BibitemOpen
      \bibfield  {author} {\bibinfo {author} {\bibfnamefont {K.}~\bibnamefont
      {Olive}} \emph {et~al.} (\bibinfo {collaboration} {Particle Data Group}),\
      }\href {http://stacks.iop.org/1674-1137/38/i=9/a=090001} {\bibfield
      {journal} {\bibinfo  {journal} {Chin. Phys. C}\ }\textbf {\bibinfo {volume}
      {38}},\ \bibinfo {pages} {090001} (\bibinfo {year} {2014})}\BibitemShut
      {NoStop}%
    \bibitem [{\citenamefont {Agostini}\ \emph {et~al.}(2018)\citenamefont
      {Agostini} \emph {et~al.}}]{Agostini:2018tnm}%
      \BibitemOpen
      \bibfield  {author} {\bibinfo {author} {\bibfnamefont {M.}~\bibnamefont
      {Agostini}} \emph {et~al.} (\bibinfo {collaboration} {GERDA}),\ }\href
      {\doibase 10.1103/PhysRevLett.120.132503} {\bibfield  {journal} {\bibinfo
      {journal} {Phys. Rev. Lett.}\ }\textbf {\bibinfo {volume} {120}},\ \bibinfo
      {pages} {132503} (\bibinfo {year} {2018})},\ \Eprint
      {http://arxiv.org/abs/1803.11100} {arXiv:1803.11100 [nucl-ex]} \BibitemShut
      {NoStop}%
    \bibitem [{\citenamefont {Aalseth}\ \emph {et~al.}(2018)\citenamefont {Aalseth}
      \emph {et~al.}}]{Aalseth:2017btx}%
      \BibitemOpen
      \bibfield  {author} {\bibinfo {author} {\bibfnamefont {C.~E.}\ \bibnamefont
      {Aalseth}} \emph {et~al.} (\bibinfo {collaboration} {Majorana}),\ }\href
      {\doibase 10.1103/PhysRevLett.120.132502} {\bibfield  {journal} {\bibinfo
      {journal} {Phys. Rev. Lett.}\ }\textbf {\bibinfo {volume} {120}},\ \bibinfo
      {pages} {132502} (\bibinfo {year} {2018})},\ \Eprint
      {http://arxiv.org/abs/1710.11608} {arXiv:1710.11608 [nucl-ex]} \BibitemShut
      {NoStop}%
    \bibitem [{\citenamefont {Wang}\ \emph {et~al.}(2017)\citenamefont {Wang} \emph
      {et~al.}}]{wang2017first}%
      \BibitemOpen
      \bibfield  {author} {\bibinfo {author} {\bibfnamefont {L.}~\bibnamefont
      {Wang}} \emph {et~al.} (\bibinfo {collaboration} {CDEX}),\ }\href
      {https://link.springer.com/article/10.1007/s11433-017-9038-4} {\bibfield
      {journal} {\bibinfo  {journal} {Sci. China-Phys. Mech. Astron.}\ }\textbf
      {\bibinfo {volume} {60}},\ \bibinfo {pages} {071011} (\bibinfo {year}
      {2017})}\BibitemShut {NoStop}%
    \bibitem [{\citenamefont {Gando}\ \emph {et~al.}(2016)\citenamefont {Gando}
      \emph {et~al.}}]{gando2016search}%
      \BibitemOpen
      \bibfield  {author} {\bibinfo {author} {\bibfnamefont {A.}~\bibnamefont
      {Gando}} \emph {et~al.} (\bibinfo {collaboration} {KamLAND-Zen}),\ }\href
      {https://journals.aps.org/prl/abstract/10.1103/PhysRevLett.117.082503}
      {\bibfield  {journal} {\bibinfo  {journal} {Phys. Rev. Lett.}\ }\textbf
      {\bibinfo {volume} {117}},\ \bibinfo {pages} {082503} (\bibinfo {year}
      {2016})}\BibitemShut {NoStop}%
    \bibitem [{\citenamefont {Albert}\ \emph {et~al.}(2018)\citenamefont {Albert}
      \emph {et~al.}}]{Albert:2017owj}%
      \BibitemOpen
      \bibfield  {author} {\bibinfo {author} {\bibfnamefont {J.~B.}\ \bibnamefont
      {Albert}} \emph {et~al.} (\bibinfo {collaboration} {EXO}),\ }\href {\doibase
      10.1103/PhysRevLett.120.072701} {\bibfield  {journal} {\bibinfo  {journal}
      {Phys. Rev. Lett.}\ }\textbf {\bibinfo {volume} {120}},\ \bibinfo {pages}
      {072701} (\bibinfo {year} {2018})},\ \Eprint
      {http://arxiv.org/abs/1707.08707} {arXiv:1707.08707 [hep-ex]} \BibitemShut
      {NoStop}%
    \bibitem [{\citenamefont {Umehara}\ \emph {et~al.}(2006)\citenamefont {Umehara}
      \emph {et~al.}}]{umehara2006candles}%
      \BibitemOpen
      \bibfield  {author} {\bibinfo {author} {\bibfnamefont {S.}~\bibnamefont
      {Umehara}} \emph {et~al.} (\bibinfo {collaboration} {CANDLES}),\ }in\ \href
      {https://iopscience.iop.org/article/10.1088/1742-6596/39/1/093/meta} {\emph
      {\bibinfo {booktitle} {Journal of Physics: Conference Series}}},\
      Vol.~\bibinfo {volume} {39}\ (\bibinfo {organization} {IOP Publishing},\
      \bibinfo {year} {2006})\ p.\ \bibinfo {pages} {356}\BibitemShut {NoStop}%
    \bibitem [{\citenamefont {Alduino}\ \emph {et~al.}(2018)\citenamefont {Alduino}
      \emph {et~al.}}]{alduino2018first}%
      \BibitemOpen
      \bibfield  {author} {\bibinfo {author} {\bibfnamefont {C.}~\bibnamefont
      {Alduino}} \emph {et~al.} (\bibinfo {collaboration} {CUORE}),\ }\href
      {https://journals.aps.org/prl/abstract/10.1103/PhysRevLett.120.132501}
      {\bibfield  {journal} {\bibinfo  {journal} {Phys. Rev. Lett.}\ }\textbf
      {\bibinfo {volume} {120}},\ \bibinfo {pages} {132501} (\bibinfo {year}
      {2018})}\BibitemShut {NoStop}%
    \bibitem [{\citenamefont {Tan}\ \emph {et~al.}(2016{\natexlab{a}})\citenamefont
      {Tan} \emph {et~al.}}]{Tan:2016diz}%
      \BibitemOpen
      \bibfield  {author} {\bibinfo {author} {\bibfnamefont {A.}~\bibnamefont
      {Tan}} \emph {et~al.} (\bibinfo {collaboration} {PandaX}),\ }\href {\doibase
      10.1103/PhysRevD.93.122009} {\bibfield  {journal} {\bibinfo  {journal} {Phys.
      Rev.}\ }\textbf {\bibinfo {volume} {D93}},\ \bibinfo {pages} {122009}
      (\bibinfo {year} {2016}{\natexlab{a}})},\ \Eprint
      {http://arxiv.org/abs/1602.06563} {arXiv:1602.06563 [hep-ex]} \BibitemShut
      {NoStop}%
    \bibitem [{\citenamefont {Tan}\ \emph {et~al.}(2016{\natexlab{b}})\citenamefont
      {Tan} \emph {et~al.}}]{tan2016dark}%
      \BibitemOpen
      \bibfield  {author} {\bibinfo {author} {\bibfnamefont {A.}~\bibnamefont
      {Tan}} \emph {et~al.} (\bibinfo {collaboration} {PandaX}),\ }\href
      {https://journals.aps.org/prl/abstract/10.1103/PhysRevLett.117.121303}
      {\bibfield  {journal} {\bibinfo  {journal} {Phys. Rev. Lett.}\ }\textbf
      {\bibinfo {volume} {117}},\ \bibinfo {pages} {121303} (\bibinfo {year}
      {2016}{\natexlab{b}})}\BibitemShut {NoStop}%
    \bibitem [{\citenamefont {Cui}\ \emph {et~al.}(2017)\citenamefont {Cui} \emph
      {et~al.}}]{cui2017dark}%
      \BibitemOpen
      \bibfield  {author} {\bibinfo {author} {\bibfnamefont {X.}~\bibnamefont
      {Cui}} \emph {et~al.} (\bibinfo {collaboration} {PandaX}),\ }\href
      {https://journals.aps.org/prl/abstract/10.1103/PhysRevLett.119.181302}
      {\bibfield  {journal} {\bibinfo  {journal} {Phys. Rev. Lett.}\ }\textbf
      {\bibinfo {volume} {119}},\ \bibinfo {pages} {181302} (\bibinfo {year}
      {2017})}\BibitemShut {NoStop}%
    \bibitem [{\citenamefont {Wu}\ \emph {et~al.}(2013)\citenamefont {Wu} \emph
      {et~al.}}]{wu2013measurement}%
      \BibitemOpen
      \bibfield  {author} {\bibinfo {author} {\bibfnamefont {Y.-C.}\ \bibnamefont
      {Wu}} \emph {et~al.},\ }\href {\doibase 10.1088/1674-1137/37/8/086001,
      10.1088/1674-1137/37/1/016001} {\bibfield  {journal} {\bibinfo  {journal}
      {Chin. Phys.}\ }\textbf {\bibinfo {volume} {C37}},\ \bibinfo {pages} {086001}
      (\bibinfo {year} {2013})},\ \Eprint {http://arxiv.org/abs/1305.0899}
      {arXiv:1305.0899 [physics.ins-det]} \BibitemShut {NoStop}%
    \bibitem [{\citenamefont {Dolgoshein}\ \emph {et~al.}(1970)\citenamefont
      {Dolgoshein}, \citenamefont {Lebedenko},\ and\ \citenamefont
      {Rodionov}}]{DualPhaseTPC}%
      \BibitemOpen
      \bibfield  {author} {\bibinfo {author} {\bibfnamefont {B.~A.}\ \bibnamefont
      {Dolgoshein}}, \bibinfo {author} {\bibfnamefont {V.~N.}\ \bibnamefont
      {Lebedenko}}, \ and\ \bibinfo {author} {\bibfnamefont {B.~U.}\ \bibnamefont
      {Rodionov}},\ }\href@noop {} {\bibfield  {journal} {\bibinfo  {journal} {JETP
      Lett.}\ }\textbf {\bibinfo {volume} {11}},\ \bibinfo {pages} {351} (\bibinfo
      {year} {1970})}\BibitemShut {NoStop}%
    \bibitem [{\citenamefont {Cao}\ \emph {et~al.}(2014)\citenamefont {Cao} \emph
      {et~al.}}]{cao2014pandax}%
      \BibitemOpen
      \bibfield  {author} {\bibinfo {author} {\bibfnamefont {X.}~\bibnamefont
      {Cao}} \emph {et~al.} (\bibinfo {collaboration} {PandaX}),\ }\href
      {https://link.springer.com/article/10.1007/s11433-014-5521-2} {\bibfield
      {journal} {\bibinfo  {journal} {Sci. China-Phys. Mech. Astron.}\ }\textbf
      {\bibinfo {volume} {57}},\ \bibinfo {pages} {1476} (\bibinfo {year}
      {2014})}\BibitemShut {NoStop}%
    \bibitem [{\citenamefont {Xiao}\ \emph {et~al.}(2015)\citenamefont {Xiao} \emph
      {et~al.}}]{Xiao:2015psa}%
      \BibitemOpen
      \bibfield  {author} {\bibinfo {author} {\bibfnamefont {X.}~\bibnamefont
      {Xiao}} \emph {et~al.} (\bibinfo {collaboration} {PandaX}),\ }\href {\doibase
      10.1103/PhysRevD.92.052004} {\bibfield  {journal} {\bibinfo  {journal} {Phys.
      Rev.}\ }\textbf {\bibinfo {volume} {D92}},\ \bibinfo {pages} {052004}
      (\bibinfo {year} {2015})},\ \Eprint {http://arxiv.org/abs/1505.00771}
      {arXiv:1505.00771 [hep-ex]} \BibitemShut {NoStop}%
    \bibitem [{\citenamefont {Yoshino}\ \emph {et~al.}(1976)\citenamefont
      {Yoshino}, \citenamefont {Sowada},\ and\ \citenamefont
      {Schmidt}}]{yoshino1976effect}%
      \BibitemOpen
      \bibfield  {author} {\bibinfo {author} {\bibfnamefont {K.}~\bibnamefont
      {Yoshino}}, \bibinfo {author} {\bibfnamefont {U.}~\bibnamefont {Sowada}}, \
      and\ \bibinfo {author} {\bibfnamefont {W.}~\bibnamefont {Schmidt}},\ }\href
      {https://journals.aps.org/pra/abstract/10.1103/PhysRevA.14.438} {\bibfield
      {journal} {\bibinfo  {journal} {Phys. Rev.}\ }\textbf {\bibinfo {volume}
      {A14}},\ \bibinfo {pages} {438} (\bibinfo {year} {1976})}\BibitemShut
      {NoStop}%
    \bibitem [{\citenamefont {{Lenardo}}\ \emph {et~al.}(2015)\citenamefont
      {{Lenardo}}, \citenamefont {{Kazkaz}}, \citenamefont {{Manalaysay}},
      \citenamefont {{Mock}}, \citenamefont {{Szydagis}},\ and\ \citenamefont
      {{Tripathi}}}]{Lenardo2015NEST}%
      \BibitemOpen
      \bibfield  {author} {\bibinfo {author} {\bibfnamefont {B.}~\bibnamefont
      {{Lenardo}}}, \bibinfo {author} {\bibfnamefont {K.}~\bibnamefont {{Kazkaz}}},
      \bibinfo {author} {\bibfnamefont {A.}~\bibnamefont {{Manalaysay}}}, \bibinfo
      {author} {\bibfnamefont {J.}~\bibnamefont {{Mock}}}, \bibinfo {author}
      {\bibfnamefont {M.}~\bibnamefont {{Szydagis}}}, \ and\ \bibinfo {author}
      {\bibfnamefont {M.}~\bibnamefont {{Tripathi}}},\ }\href {\doibase
      10.1109/TNS.2015.2481322} {\bibfield  {journal} {\bibinfo  {journal} {IEEE
      Trans. Nucl. Sci.}\ }\textbf {\bibinfo {volume} {62}},\ \bibinfo {pages}
      {3387} (\bibinfo {year} {2015})}\BibitemShut {NoStop}%
    \bibitem [{\citenamefont {Agostinelli}\ \emph {et~al.}(2003)\citenamefont
      {Agostinelli} \emph {et~al.}}]{agostinelli2003geant4}%
      \BibitemOpen
      \bibfield  {author} {\bibinfo {author} {\bibfnamefont {S.}~\bibnamefont
      {Agostinelli}} \emph {et~al.} (\bibinfo {collaboration} {GEANT4}),\ }\href
      {https://www.sciencedirect.com/science/article/pii/S0168900203013688}
      {\bibfield  {journal} {\bibinfo  {journal} {Nucl. Instrum. Meth.}\ }\textbf
      {\bibinfo {volume} {A506}},\ \bibinfo {pages} {250} (\bibinfo {year}
      {2003})}\BibitemShut {NoStop}%
    \bibitem [{\citenamefont {Wang}\ \emph {et~al.}(2016)\citenamefont {Wang},
      \citenamefont {Chen}, \citenamefont {Fu}, \citenamefont {Ji}, \citenamefont
      {Liu}, \citenamefont {Mao}, \citenamefont {Wang}, \citenamefont {Wang},
      \citenamefont {Xie},\ and\ \citenamefont {Zhang}}]{Wang2016counting}%
      \BibitemOpen
      \bibfield  {author} {\bibinfo {author} {\bibfnamefont {X.}~\bibnamefont
      {Wang}}, \bibinfo {author} {\bibfnamefont {X.}~\bibnamefont {Chen}}, \bibinfo
      {author} {\bibfnamefont {C.}~\bibnamefont {Fu}}, \bibinfo {author}
      {\bibfnamefont {X.}~\bibnamefont {Ji}}, \bibinfo {author} {\bibfnamefont
      {X.}~\bibnamefont {Liu}}, \bibinfo {author} {\bibfnamefont {Y.}~\bibnamefont
      {Mao}}, \bibinfo {author} {\bibfnamefont {H.}~\bibnamefont {Wang}}, \bibinfo
      {author} {\bibfnamefont {S.}~\bibnamefont {Wang}}, \bibinfo {author}
      {\bibfnamefont {P.}~\bibnamefont {Xie}}, \ and\ \bibinfo {author}
      {\bibfnamefont {T.}~\bibnamefont {Zhang}},\ }\href {\doibase
      10.1088/1748-0221/11/12/t12002} {\bibfield  {journal} {\bibinfo  {journal}
      {J. Instrum.}\ }\textbf {\bibinfo {volume} {11}},\ \bibinfo {pages} {T12002}
      (\bibinfo {year} {2016})}\BibitemShut {NoStop}%
    \bibitem [{\citenamefont {Li}\ \emph {et~al.}(2017)\citenamefont {Li} \emph
      {et~al.}}]{li2017krypton}%
      \BibitemOpen
      \bibfield  {author} {\bibinfo {author} {\bibfnamefont {S.}~\bibnamefont {Li}}
      \emph {et~al.} (\bibinfo {collaboration} {PandaX}),\ }\href
      {https://iopscience.iop.org/article/10.1088/1748-0221/12/02/T02002/pdf}
      {\bibfield  {journal} {\bibinfo  {journal} {J. Instrum.}\ }\textbf {\bibinfo
      {volume} {12}},\ \bibinfo {pages} {T02002} (\bibinfo {year}
      {2017})}\BibitemShut {NoStop}%
    \bibitem [{\citenamefont {Ponkratenko}\ \emph {et~al.}(2000)\citenamefont
      {Ponkratenko}, \citenamefont {Tretyak},\ and\ \citenamefont
      {Zdesenko}}]{ponkratenko2000event}%
      \BibitemOpen
      \bibfield  {author} {\bibinfo {author} {\bibfnamefont {O.}~\bibnamefont
      {Ponkratenko}}, \bibinfo {author} {\bibfnamefont {V.}~\bibnamefont
      {Tretyak}}, \ and\ \bibinfo {author} {\bibfnamefont {Y.~G.}\ \bibnamefont
      {Zdesenko}},\ }\href {https://link.springer.com/article/10.1134/1.855784}
      {\bibfield  {journal} {\bibinfo  {journal} {Phys. Atom. Nuclei}\ }\textbf
      {\bibinfo {volume} {63}},\ \bibinfo {pages} {1282} (\bibinfo {year}
      {2000})}\BibitemShut {NoStop}%
    \bibitem [{\citenamefont {Albert}\ \emph {et~al.}(2014)\citenamefont {Albert}
      \emph {et~al.}}]{Albert:2013gpz}%
      \BibitemOpen
      \bibfield  {author} {\bibinfo {author} {\bibfnamefont {J.~B.}\ \bibnamefont
      {Albert}} \emph {et~al.} (\bibinfo {collaboration} {EXO-200}),\ }\href
      {\doibase 10.1103/PhysRevC.89.015502} {\bibfield  {journal} {\bibinfo
      {journal} {Phys. Rev.}\ }\textbf {\bibinfo {volume} {C89}},\ \bibinfo {pages}
      {015502} (\bibinfo {year} {2014})},\ \Eprint {http://arxiv.org/abs/1306.6106}
      {arXiv:1306.6106 [nucl-ex]} \BibitemShut {NoStop}%
    \bibitem [{\citenamefont {Caldwell}\ \emph {et~al.}(2009)\citenamefont
      {Caldwell}, \citenamefont {Koll{\'a}r},\ and\ \citenamefont
      {Kr{\"o}ninger}}]{caldwell2009bat}%
      \BibitemOpen
      \bibfield  {author} {\bibinfo {author} {\bibfnamefont {A.}~\bibnamefont
      {Caldwell}}, \bibinfo {author} {\bibfnamefont {D.}~\bibnamefont
      {Koll{\'a}r}}, \ and\ \bibinfo {author} {\bibfnamefont {K.}~\bibnamefont
      {Kr{\"o}ninger}},\ }\href
      {https://www.sciencedirect.com/science/article/pii/S0010465509002045}
      {\bibfield  {journal} {\bibinfo  {journal} {Comput. Phys. Commun.}\ }\textbf
      {\bibinfo {volume} {180}},\ \bibinfo {pages} {2197} (\bibinfo {year}
      {2009})}\BibitemShut {NoStop}%
    \bibitem [{\citenamefont {Alduino}\ \emph {et~al.}(2016)\citenamefont {Alduino}
      \emph {et~al.}}]{alduino2016analysis}%
      \BibitemOpen
      \bibfield  {author} {\bibinfo {author} {\bibfnamefont {C.}~\bibnamefont
      {Alduino}} \emph {et~al.} (\bibinfo {collaboration} {CUORE}),\ }\href
      {https://journals.aps.org/prc/abstract/10.1103/PhysRevC.93.045503} {\bibfield
       {journal} {\bibinfo  {journal} {Phys. Rev.}\ }\textbf {\bibinfo {volume}
      {C93}},\ \bibinfo {pages} {045503} (\bibinfo {year} {2016})}\BibitemShut
      {NoStop}%
    \bibitem [{\citenamefont {Casella}\ and\ \citenamefont
      {Berger}(2002)}]{casella2002statistical}%
      \BibitemOpen
      \bibfield  {author} {\bibinfo {author} {\bibfnamefont {G.}~\bibnamefont
      {Casella}}\ and\ \bibinfo {author} {\bibfnamefont {R.~L.}\ \bibnamefont
      {Berger}},\ }\href@noop {} {\bibfield  {journal} {\bibinfo  {journal}
      {Pacific Grove, CA}\ } (\bibinfo {year} {2002})}\BibitemShut {NoStop}%
    \bibitem [{\citenamefont {Kotila}\ and\ \citenamefont
      {Iachello}(2012)}]{kotila2012phase}%
      \BibitemOpen
      \bibfield  {author} {\bibinfo {author} {\bibfnamefont {J.}~\bibnamefont
      {Kotila}}\ and\ \bibinfo {author} {\bibfnamefont {F.}~\bibnamefont
      {Iachello}},\ }\href
      {https://journals.aps.org/prc/abstract/10.1103/PhysRevC.85.034316} {\bibfield
       {journal} {\bibinfo  {journal} {Phys. Rev.}\ }\textbf {\bibinfo {volume}
      {C85}},\ \bibinfo {pages} {034316} (\bibinfo {year} {2012})}\BibitemShut
      {NoStop}%
    \bibitem [{\citenamefont {Barea}\ \emph {et~al.}(2015)\citenamefont {Barea},
      \citenamefont {Kotila},\ and\ \citenamefont {Iachello}}]{barea20150}%
      \BibitemOpen
      \bibfield  {author} {\bibinfo {author} {\bibfnamefont {J.}~\bibnamefont
      {Barea}}, \bibinfo {author} {\bibfnamefont {J.}~\bibnamefont {Kotila}}, \
      and\ \bibinfo {author} {\bibfnamefont {F.}~\bibnamefont {Iachello}},\ }\href
      {https://journals.aps.org/prc/abstract/10.1103/PhysRevC.91.034304} {\bibfield
       {journal} {\bibinfo  {journal} {Phys. Rev.}\ }\textbf {\bibinfo {volume}
      {C91}},\ \bibinfo {pages} {034304} (\bibinfo {year} {2015})}\BibitemShut
      {NoStop}%
    \bibitem [{\citenamefont {Rodr{\'\i}guez}\ and\ \citenamefont
      {Mart{\'\i}nez-Pinedo}(2010)}]{rodriguez2010energy}%
      \BibitemOpen
      \bibfield  {author} {\bibinfo {author} {\bibfnamefont {T.~R.}\ \bibnamefont
      {Rodr{\'\i}guez}}\ and\ \bibinfo {author} {\bibfnamefont {G.}~\bibnamefont
      {Mart{\'\i}nez-Pinedo}},\ }\href
      {https://journals.aps.org/prl/abstract/10.1103/PhysRevLett.105.252503}
      {\bibfield  {journal} {\bibinfo  {journal} {Phys. Rev. Lett.}\ }\textbf
      {\bibinfo {volume} {105}},\ \bibinfo {pages} {252503} (\bibinfo {year}
      {2010})}\BibitemShut {NoStop}%
    \bibitem [{\citenamefont {Engel}\ \emph {et~al.}(2014)\citenamefont {Engel},
      \citenamefont {{\v{S}}imkovic},\ and\ \citenamefont
      {Vogel}}]{engel2014chiral}%
      \BibitemOpen
      \bibfield  {author} {\bibinfo {author} {\bibfnamefont {J.}~\bibnamefont
      {Engel}}, \bibinfo {author} {\bibfnamefont {F.}~\bibnamefont
      {{\v{S}}imkovic}}, \ and\ \bibinfo {author} {\bibfnamefont {P.}~\bibnamefont
      {Vogel}},\ }\href
      {https://journals.aps.org/prc/abstract/10.1103/PhysRevC.89.064308} {\bibfield
       {journal} {\bibinfo  {journal} {Phys. Rev.}\ }\textbf {\bibinfo {volume}
      {C89}},\ \bibinfo {pages} {064308} (\bibinfo {year} {2014})}\BibitemShut
      {NoStop}%
    \bibitem [{\citenamefont {Menendez}\ \emph {et~al.}(2009)\citenamefont
      {Menendez}, \citenamefont {Poves}, \citenamefont {Caurier},\ and\
      \citenamefont {Nowacki}}]{menendez2009disassembling}%
      \BibitemOpen
      \bibfield  {author} {\bibinfo {author} {\bibfnamefont {J.}~\bibnamefont
      {Menendez}}, \bibinfo {author} {\bibfnamefont {A.}~\bibnamefont {Poves}},
      \bibinfo {author} {\bibfnamefont {E.}~\bibnamefont {Caurier}}, \ and\
      \bibinfo {author} {\bibfnamefont {F.}~\bibnamefont {Nowacki}},\ }\href
      {https://www.sciencedirect.com/science/article/pii/S0375947408008233}
      {\bibfield  {journal} {\bibinfo  {journal} {Nucl. Phys.}\ }\textbf {\bibinfo
      {volume} {A818}},\ \bibinfo {pages} {139} (\bibinfo {year}
      {2009})}\BibitemShut {NoStop}%
    \bibitem [{\citenamefont {Mustonen}\ and\ \citenamefont
      {Engel}(2013)}]{mustonen2013large}%
      \BibitemOpen
      \bibfield  {author} {\bibinfo {author} {\bibfnamefont {M.}~\bibnamefont
      {Mustonen}}\ and\ \bibinfo {author} {\bibfnamefont {J.}~\bibnamefont
      {Engel}},\ }\href
      {https://journals.aps.org/prc/abstract/10.1103/PhysRevC.87.064302} {\bibfield
       {journal} {\bibinfo  {journal} {Phys. Rev.}\ }\textbf {\bibinfo {volume}
      {C87}},\ \bibinfo {pages} {064302} (\bibinfo {year} {2013})}\BibitemShut
      {NoStop}%
    \bibitem [{\citenamefont {Aprile}\ \emph {et~al.}(2019)\citenamefont {Aprile}
      \emph {et~al.}}]{XENON:2019dti}%
      \BibitemOpen
      \bibfield  {author} {\bibinfo {author} {\bibfnamefont {E.}~\bibnamefont
      {Aprile}} \emph {et~al.} (\bibinfo {collaboration} {XENON}),\ }\href
      {\doibase 10.1038/s41586-019-1124-4} {\bibfield  {journal} {\bibinfo
      {journal} {Nature}\ }\textbf {\bibinfo {volume} {568}},\ \bibinfo {pages}
      {532} (\bibinfo {year} {2019})},\ \Eprint {http://arxiv.org/abs/1904.11002}
      {arXiv:1904.11002 [nucl-ex]} \BibitemShut {NoStop}%
    \bibitem [{\citenamefont {Zhang}\ \emph {et~al.}(2019)\citenamefont {Zhang}
      \emph {et~al.}}]{zhang2019dark}%
      \BibitemOpen
      \bibfield  {author} {\bibinfo {author} {\bibfnamefont {H.}~\bibnamefont
      {Zhang}} \emph {et~al.} (\bibinfo {collaboration} {PandaX}),\ }\href
      {https://link.springer.com/article/10.1007/s11433-018-9259-0} {\bibfield
      {journal} {\bibinfo  {journal} {Sci. China-Phys. Mech. Astron.}\ }\textbf
      {\bibinfo {volume} {62}},\ \bibinfo {pages} {31011} (\bibinfo {year}
      {2019})}\BibitemShut {NoStop}%
    \bibitem [{\citenamefont {Baudis}\ \emph {et~al.}(2014)\citenamefont {Baudis},
      \citenamefont {Ferella}, \citenamefont {Kish}, \citenamefont {Manalaysay},
      \citenamefont {Marrodan~Undagoitia},\ and\ \citenamefont
      {Schumann}}]{Baudis:2013qla}%
      \BibitemOpen
      \bibfield  {author} {\bibinfo {author} {\bibfnamefont {L.}~\bibnamefont
      {Baudis}}, \bibinfo {author} {\bibfnamefont {A.}~\bibnamefont {Ferella}},
      \bibinfo {author} {\bibfnamefont {A.}~\bibnamefont {Kish}}, \bibinfo {author}
      {\bibfnamefont {A.}~\bibnamefont {Manalaysay}}, \bibinfo {author}
      {\bibfnamefont {T.}~\bibnamefont {Marrodan~Undagoitia}}, \ and\ \bibinfo
      {author} {\bibfnamefont {M.}~\bibnamefont {Schumann}},\ }\href {\doibase
      10.1088/1475-7516/2014/01/044} {\bibfield  {journal} {\bibinfo  {journal}
      {JCAP}\ }\textbf {\bibinfo {volume} {1401}},\ \bibinfo {pages} {044}
      (\bibinfo {year} {2014})},\ \Eprint {http://arxiv.org/abs/1309.7024}
      {arXiv:1309.7024 [physics.ins-det]} \BibitemShut {NoStop}%
    \end{thebibliography}
\end{document}